\documentclass[twocolumn]{aastex631}

\newcommand{\kms}{$\,$km$\,$s$^{-1}$}

\newcommand{\target}{KSP-OT-201712a}
\newcommand{\targetold}{KSP-OT-201701a}

\newcommand{\ha}{\hbox{H$\alpha$}}
\newcommand{\hb}{\hbox{H$\beta$}}
\newcommand{\hei}{\hbox{He\,{\sc i}}}

\newcommand{\nad}{\hbox{Na\,D}}
\newcommand{\nai}{\hbox{Na\,I}}
\newcommand{\mgb}{\hbox{Mg\,$b$}}
\newcommand\vi{\mbox{$V\!-\!I$}}
\newcommand{\zc}{Z~Cam}

\newcommand{\ug}{U~Gem}

\newcommand{\su}{SU~UMa}

\newcommand{\amcvn}{AM~CVn}

\begin{document}

\title{
Helium-deficient ER UMa-type dwarf nova below the period minimum with a hot secondary
}
\shorttitle{Lee et al.}

\correspondingauthor{Youngdae Lee}
\email{hippo206@gmail.com}

\author[0000-0002-6261-1531]{Youngdae Lee} 
\affil{Department of Astronomy and Space Science, Chungnam National University, Daejeon 34134, Republic of Korea}
\affil{Research Institute of Natural Sciences, Chungnam National University, Daejeon 34134, Republic of Korea}

\author[0000-0003-4200-5064]{Dae-Sik Moon} 
\affil{David A. Dunlap Department of Astronomy and Astrophysics, University of Toronto, 50 St. George Street, Toronto, ON M5S 3H4, Canada}

\author[0000-0001-9670-1546]{Sang Chul Kim} 
\affil{Korea Astronomy and Space Science Institute 776, Daedeokdae-ro, Yuseong-gu, Daejeon 34055, Republic of Korea}
\affil{Korea National University of Science \& Technology (UST), 217 Gajeong-ro, Yuseong-gu, Daejeon 34113, Republic of Korea}

\author[0000-0002-3505-3036]{Hong Soo Park} 
\affil{Korea Astronomy and Space Science Institute 776, Daedeokdae-ro, Yuseong-gu, Daejeon 34055, Republic of Korea}
\affil{Korea National University of Science \& Technology (UST), 217 Gajeong-ro, Yuseong-gu, Daejeon 34113, Republic of Korea}

\author[0000-0003-3656-5268]{Yuan Qi Ni} 
\affil{David A. Dunlap Department of Astronomy and Astrophysics, University of Toronto, 50 St. George Street, Toronto, ON M5S 3H4, Canada}

\shorttitle{Dwarf nova : {\target}}
\shortauthors{Lee et al.}

\begin{abstract}
We present the discovery of a peculiar dwarf nova \target\ using high-cadence, multi-color
observations made with the Korea Microlensing Telescope Network.
{\target} exhibits a rare presence of outbursts during standstills 
as well as strong \ha\ emission for a dwarf nova 
below the period minimum with an orbital
period of $58.75 \pm 0.02$ minutes. 
The outburst cycles are $\sim$ 6.6 days within standstills
but increase to $\sim$ 15 days outside of them. 
Both \bv\ and \vi\ colors become bluer and redder as the outburst luminosities
increase and decrease, respectively, for the outburst within standstill,
while they evolve in the opposite directions 
outside of the standstills.
The presence of strong double-peaked {\ha} and weak {\hei} emission lines with He/H flux ratio of 0.27, together with 
absorption lines of {\mgb} and {\nad} in the source,
leads to the estimation ${\rm T}_{\rm eff}$ $\simeq$ $4570 \pm 40$~K, 
${\rm [Fe/H]}$ $\simeq$  $0.06 \pm 0.15$~dex, and log~$g$ $\simeq$ $4.5 \pm 0.1$ for its secondary.
\target\ is the second He-deficient dwarf nova below the period minimum,
while the temperature of the secondary is measured for the first time in such objects.
We identify it to be an ER UMa type dwarf nova
suggesting that the evolution of dwarf novae across 
the period minimum is accompanied by large mass 
transfers.
The high temperature of the secondary 
indicates that the system started its mass transfer 
when the secondary was about 93\% of its main sequence age.
The system will evolve to a helium cataclysmic variable or to \amcvn\ once its hydrogen envelope is exhausted
before it explodes as a Type Ia supernova.

\end{abstract}

\keywords{stars: dwarf novae --- surveys --- techniques: photometric, spectroscopic}

\section{Introduction} \label{sec:intro}

Dwarf novae are low-mass binaries composed of a white dwarf (WD)
and a secondary mass donor together with an accretion disk \citep{War95}.
Their orbital periods are $\lesssim$10 hours \citep{Rit03}, indicating that 
they are at a late dynamical evolutionary stage of binary systems
and their orbits shrink due to gravitational radiation or magnetic braking
as they evolve until the orbital periods become close to the period minimum about 75 minutes \citep{Pac81,Rap83,Kni11,Gol15}.
Some dwarf novae are expected to evolve to AM Canum Venaticorum ({\amcvn}) objects \citep{Ken15,Bur22,Bel23},
which are WD + WD binaries with an extremely short ($\lesssim$ 40~minutes) orbital period.
\amcvn\ systems are considered to be progenitors of some Type Ia supernovae 
as well as sources of substantial gravitational wave radiation \citep{Han04,Bil07,ElB21}.
It is believed that dwarf novae with a high central density of the secondary star 
at the onset of mass transfer evolve to {\amcvn} 
than those with a low central density. 
According to \citet{Pod03}, low-mass binaries with a central hydrogen abundance in mass, $X_c$,
smaller than  0.4 can evolve to {\amcvn} systems.
Dwarf novae with a more evolved secondary star 
typically show a relatively large helium to hydrogen abundance ratio at the center of the secondary star.

Dwarf nova outbursts are triggered by thermal instability in the accretion disk,
with their properties determined by the mass transfer from the secondary star
and binary properties \citep{Can87,Osa96}.
Dwarf novae are generally categorized into three groups based on 
the observed outburst properties: 
\ug\ (Geminorum), \su\ (Ursae Majoris), and \zc\ (Camelopardalis) types.
About 13\% of dwarf novae are \ug\ type that shows both short ($\sim$ 5 days) and long ($\sim$ 12 days) outbursts,
and their orbital periods are typically larger than the period gap of 2--3 hours \citep{Sma00,Otu16}.
\su\ types amount to about 70\% of the entire dwarf nova population showing the presence of normal and superoutbursts 
with their orbital periods smaller than the period gap \citep{Otu16}.
The short outbursts from \ug\ and normal outbursts from \su\ are almost identical.
There exist small differences between the long and superoutbursts
in that the latter contain superhumps of magnitude variation of about 0.2 mag
and plateaus during their decline phase \citep{War95}.

\zc\ types are very rare (about 5\%) compared to \ug\ and \su\ types.
Their outburst light curves are similar to those of \ug\ types but
featured with post-outburst ``standstills" when 
the mass-transfer rate approaches the critical value $\dot{M}_{\rm crit} \sim 8\times10^{17}\,{\rm g}\,{\rm s}^{-1}$ depending on disk radius, viscosity, orbital period and mass ratio of the system \citep{Mey83,Otu16,Bua01b}.
Although standstills in \zc\ types typically show very small magnitude variations, 
there has been a growing number of \zc\ types that show the presence of
outbursts towards the end of standstills \citep{Sim11,Szk13,Kat19a,Kat19c}.
The orbital periods of \zc\ types are known to be larger than the period gap 
as \ug\ types \citep{Otu16}.
One notable exception is the peculiar case of NY Serpentis (NY Ser), 
a dwarf nova which shows characteristics of both \su\ type-like superoutbursts 
and \zc\ type-like standstills together \citep{Kat19a}.
ER UMa types also show standstills and large mass-transfer rate \citep[MGAB-V859, ZTF18abgjsdg, BO Cet;][] {Kat21a,Kat21b}, appearing to be Z Cam types below the period gap.

Only a limited number of dwarf novae with an orbital period smaller than
the so-called period minimum of $\sim$ 75 minutes \citep{Bre12,Ken15}, have been found:
EI Psc \citep{Tho02a}, ZTF J1813+4251 \citep{Bur22}, KSP-OT-201701a \citep{Lee22}, OV Boo \citep{Pat08}, V485 Cen \citep{Aug96,Ole97}, CSS120422 \citep{Car13}, V418 Ser \citep{Ken15}, and CSS100603 \citep{Bre12}.
Most of these objects are helium dwarf novae with strong helium and hydrogen emission lines and are believed to be in the evolutionary path to \amcvn. 
In the case of KSP-OT-201701a, the helium to hydrogen flux ratio
is notably smaller than those of the other objects, indicating that 
it is in an earlier evolutionary stage to \amcvn\ than the others \citep{Lee22},
although the properties of its secondary star remain largely unknown. 
Because of the lack of observed examples, however, the details of
how short-period dwarf novae evolve toward the \amcvn\ phase are poorly understood.

In this paper, we provide an illustrative case of \target\
that sheds new light on the early evolutionary stage of 
a short-period dwarf nova to an {\amcvn}, revealing the properties of the secondary.

\section{{\target}: Discovery and Light Curves} \label{sec:discovery}

As part of the Korea Microlensing Telescope Network (KMTNet) Supernova Program \citep[KSP;][]{Moo16} aimed
at detecting infant/early supernovae \citep{Afs19,Moo21,Ni22,Ni23b,Ni23}, 
we monitored the 4-square degree field around the lenticular galaxy NGC 2380 
in the $BVI$ bands with an average cadence of 8 hours for each band 
between December 2017 and March 2019 using the three 1.6-m telescopes of 
the KMTNet located at Chile, South Africa and Australia \citep{Kim16}.
All our images are 60 sec exposures reaching an average detection limit of 21.5 mag.
In addition to infant/early supernovae, KSP has also discovered other types of optical transients, including several novae and dwarf novae of interest \citep{Ant17,Bro18,Lee19,Lee22}.

We discovered a dwarf nova \target\ at the coordinate  
$(\alpha,\delta)_{J2000}$ = (111.0321\degr, --26.3318\degr)\footnote{$(l,b)$ = (239.9753\degr, --5.0985\degr)} on December 25, 2017 (UT).
This object was also observed by the Wide-field Infrared Survey Explorer (WISE) with a single exposure \citep{Wri10} and the Zwicky Transient Facility (ZTF) with low- and high-cadence time-series observations in $r$ band for about 2 years from 2018 November \citep{Bel19}.
We performed DAOPHOT-based point spread function photometry \citep[using DAOPHOT II;][]{Ste87}.
For the photometric calibration of our KSP $BVI$ images, we used about 300 $BVi$ standard stars near the source in the AAVSO Photometric All Sky Survey (APASS) database and made $I=i^\prime - 0.4$ mag \citep{Par17,Lee19}.
$B$-band KMTNet instrumental magnitudes required a color correction and were corrected by $\simeq 0.27\times(B-V)$ \citep[see ][for details]{Par17}.
Extinction corrections $A_B = 1.57$ mag, $A_V = 1.18$ mag, and $A_I = 0.72$ mag were applied (see Section~\ref{sec:extinc}).
Throughout this paper, we use extinction-corrected magnitudes and colors unless otherwise specified.

\begin{figure*}
\epsscale{1.2}
\plotone{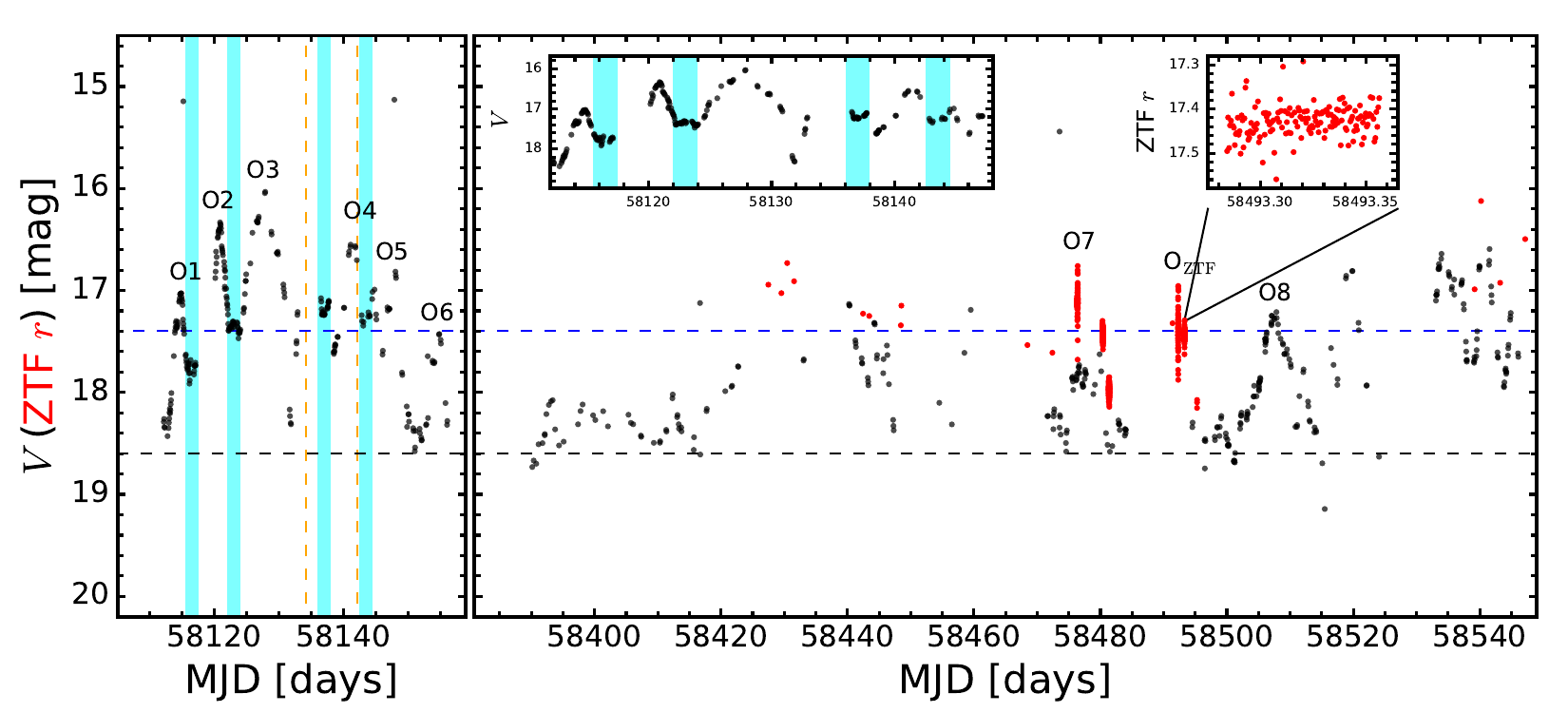}
\caption{The observed $V$- and $r$-band light curves of {\target} 
by the KMTNet and ZTF are shown in filled black and red circles, respectively.
The panel on the left-hand side is for the period of MJD = 58110--58160,
whereas the panel on the right-hand side is for MJD = 58380--58550.
On the left-hand panel, we identify six outbursts 
that show a clear rise and decay across peak brightness 
which we name O1--O6 from left to right.
On the right-hand panel, we identify at least two such outbursts
of O7 and O8 from the KMTNet data. 
Although there exist more outburst activities,
only these two outbursts are clearly 
identified by observations of their peak brightnesses.
During the period of MJD = 58475--58495, ZTF also detected
outburst activities of the source in the $r$ band (filled red circles), 
consisting of four separate 2-hour and one 40-minute (MJD $\sim$ 58490 day) observations
conducted at about 40-second cadence and showing an $r$-band
magnitude variation of 18.2--16.8 mag. 
The first three 2-hour ZTF observations overlap with 
the O7 outburst identified by the KMTNet data.
The fourth 2-hour observation occurred at MJD $\sim$ 58493.30 day
when there was no KMTNet observation,
and we named this O$_{\rm ZTF}$.
The inset in the right-hand side shows how the $r$-band magnitude changes
during O$_{\rm ZTF}$.
The two horizontal dashed lines represent the quiescent magnitude
($V$ = 18.6 mag; black) of the source and
$V$ = 17.4 mag (blue) which is the brightness of the 
standstill just after the O2 outburst.
The shaded cyan vertical area on the left-hand plot 
corresponds to the interval of the standstills therein.
The inset in the left-hand side provides an enlarged view of part of the light curve to highlight the shape of the standstills.
The two dashed orange vertical lines mark the epochs 
of spectroscopic observations of the source. 
Extinction is not corrected.
Magnitude errors are about 0.02 mag around $V = 18.6$ mag. 
}
\label{fulllc}
\end{figure*}

\begin{figure*}

\epsscale{1.0}
\plotone{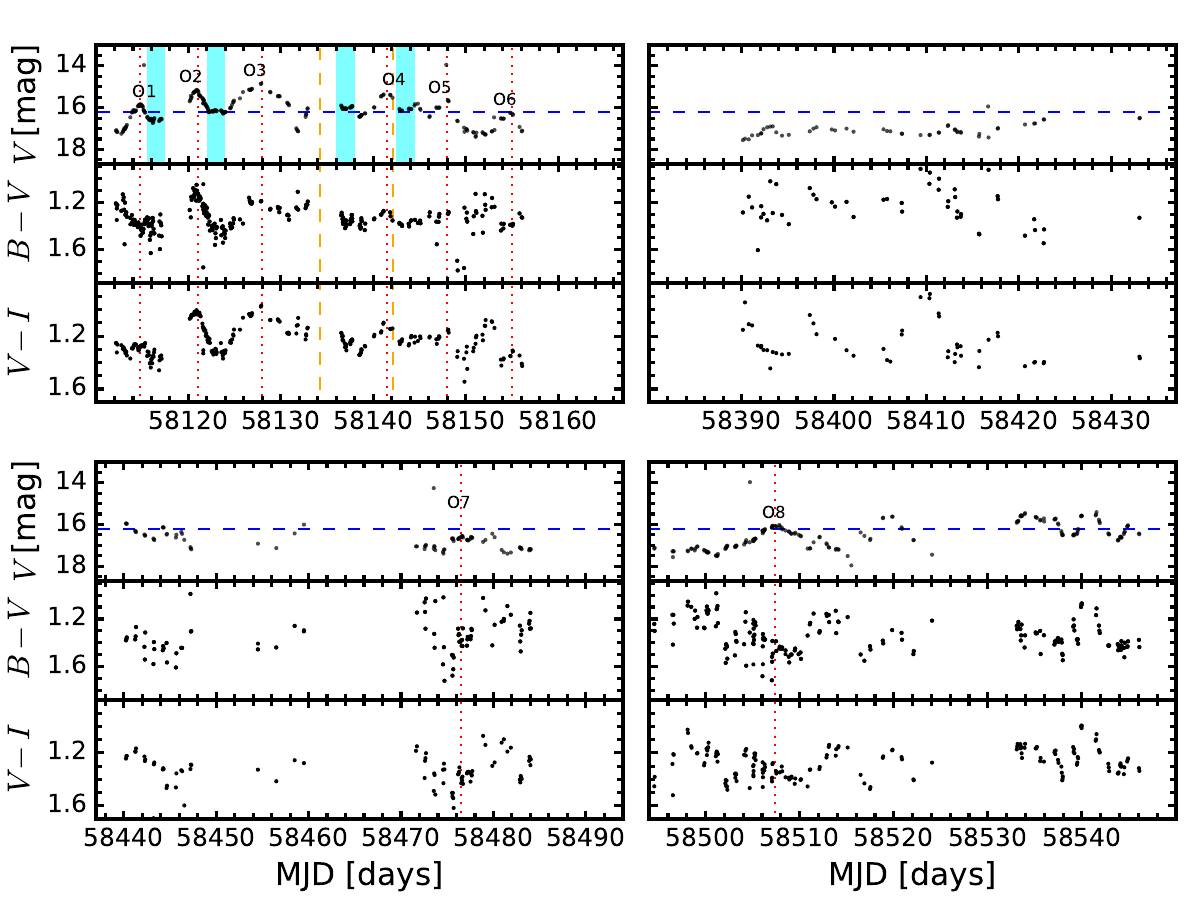}
\caption{(Top-left panel) The \bv\ (middle plot) and \vi\ (bottom
plot) color evolution of \target\ synchronous with that of 
the $V$-band light curve (top plot) during the period 
of MJD = 58110--58166 which corresponds to the left-hand panel in Figure~\ref{fulllc}.
All magnitudes are extinction corrected.
(Top-right panel) Same as the top-left panel, but for 
the first one-third period (MJD = 58380--58437 days) 
of the right panel in Figure~\ref{fulllc}. 
(Bottom-left panel) Same as the top-right panel but for 
the second one-third period (MJD = 58437--58494 days). 
(Bottom-right panel) Same as the top-right panel, 
but for the final one-third period (MJD = 58494--58550 days).
The shaded cyan vertical areas and the blue horizontal dashed lines are the same in Figure~\ref{fulllc}.
The red vertical dotted lines indicate the peak epochs of the O1--O8 outbursts.
}
\label{zoomlc}
\end{figure*}

Figure~\ref{fulllc} shows the extinction-uncorrected $V$-band light curves of {\target} revealing the abundance of outburst activities from the source.
In the left panel, we identify at least six outbursts that we name 
O1-O6 between MJD 58110 and 58160.
We summarize the observed properties of them as follows:
(1) The six outbursts are overlapped in which the next outburst starts
before the previous one reaches the quiescent brightness (see below).
(2) The peak brightness of each outburst increases until O3 after which it decreases,
and the average outburst cycle is about 6.6 days.
(3) Their rising rates are 0.79, 0.57, 0.32, 0.43, 0.44 and 0.36 mag day$^{-1}$ for O1-O6, respectively, with the primary peak O3 having the smallest value.
(4) The amplitudes of these outbursts, which are defined as the magnitude difference
between the start of outburst rise and peak, 
are 1.52, 1.47, 1.34, 1.22, 0.85, and 1.02 mag for O1-O6, respectively. 
(5) The intermediate phases, which we identify to be ``standstills" (\S\ref{sec:ori}), showing almost constant brightness 
occur between overlapped outbursts (see cyan shaded areas in Figure~\ref{fulllc} and \ref{zoomlc}) 
and typically last about 2 days. 
The brightness of the intermediate phase after O2 is shown as a guideline (blue dashed line).
(6) The minimum brightness of the object is 18.6 mag, or 17.42 mag with extinction correction (\S\ref{sec:extinc}),
which we adopt as quiescent magnitude shown as the black dashed line.

In the right panel of Figure~\ref{fulllc} during the period of MJD 58380--58550, 
the source is featured with complicated, 
but different from those in the left panel, outburst activities
with varying amplitudes ranging from 0.5 mag to $\gtrsim$ 2 mag.
Although the lack of observations in some intervals during the period
makes it difficult to clearly identify outburst peaks, 
we identify at least two peaks which we name O7 and O8 (black dots).
In particular, the O8 outburst, which peaks at MJD 58507.3748,
occurs between two quiescent phases
with peak amplitude and rising rate of 1.60 mag and 0.28 mag day$^{-1}$, respectively.
Using the O8 outburst peak, we estimate the distance to the source 
to be about 1.32 kpc (see \S~\ref{sec:extinc}).
The red dots in Figure~\ref{fulllc} are the $r$-band magnitudes of the
source observed by the Zwicky Transient Facility \citep[ZTF;][]{Bel19}.
The ZTF observations are both with high (40 seconds) and low ($>$ 1 day) cadences with
part of them covering the intervals when there are no KMTNet observations.
The outburst (O$_{\rm ZTF}$) around MJD 58493 is largely detected by ZTF solely
since the KMTNet observations were made for only its tail phase.
The subset in Figure~\ref{fulllc} shows the ZTF light curve 
of O$_{\rm ZTF}$ in which the source magnitudes increase at 
a rate of 0.37 mag day$^{-1}$ as typically found in dwarf nova outbursts \citep[e.g.][]{Otu16}.
The outburst cycle between O$_{\rm ZTF}$ and O8,
where there are no significant observational gaps from the combined KMTNet and ZTF observations,
is about 15 days,
which is equivalent to the cycle between the O7 and O$_{\rm ZTF}$ outbursts.

Figure~\ref{zoomlc} shows the extinction-corrected color evolution 
of \target\ in which its \bv\ and \vi\ colors vary in a similar manner.
In the overlapping O2--O5 outbursts,
the colors become bluest at their peak brightness 
as typically found in outbursts of other dwarf novae \citep{Can87}.
On the contrary,  the colors of O7 and O8, as well as those of O1 and O6, 
show an almost opposite behavior with redward evolutions towards the peak brightness,
attaining maximum redness a few days after the peak.

\section{Spectroscopy} \label{sec:spec}
Spectroscopic observations of \target\ were made
with Gemini Multi-object Spectrographs South (GMOS-S) at two epochs
of 2018 January 16 (UT, MJD 58134.2016) and 24 (UT, MJD 58142.1823).
Four 400-sec exposures were obtained for the blue channel  (3970--7060${\rm \AA}$)
with a 1.5$\arcsec$ width slit during the first epoch, 
while four 280-sec exposures were obtained for 
each of the blue channel and red channel (5405--10000${\rm \AA}$) during the second
epoch.
The spectral resolutions measured in FWHM were 2.73 ${\rm \AA}$ at $4610 {\rm \AA}$ and
3.98 ${\rm \AA}$ at $7640 {\rm \AA}$ for the blue and red channels, respectively.
The Gemini IRAF\footnote{IRAF is distributed by National Optical Astronomy Observatories, which is operated by the Association of Universities for Research in Astronomy, Inc. (AURA), under cooperative agreement with the National Science Foundation, USA.} package was used to conduct basic data reduction
of the observed spectra, including image preprocessing, wavelength calibration, and flux calibration.

\begin{figure*}
\epsscale{1.0}
\plotone{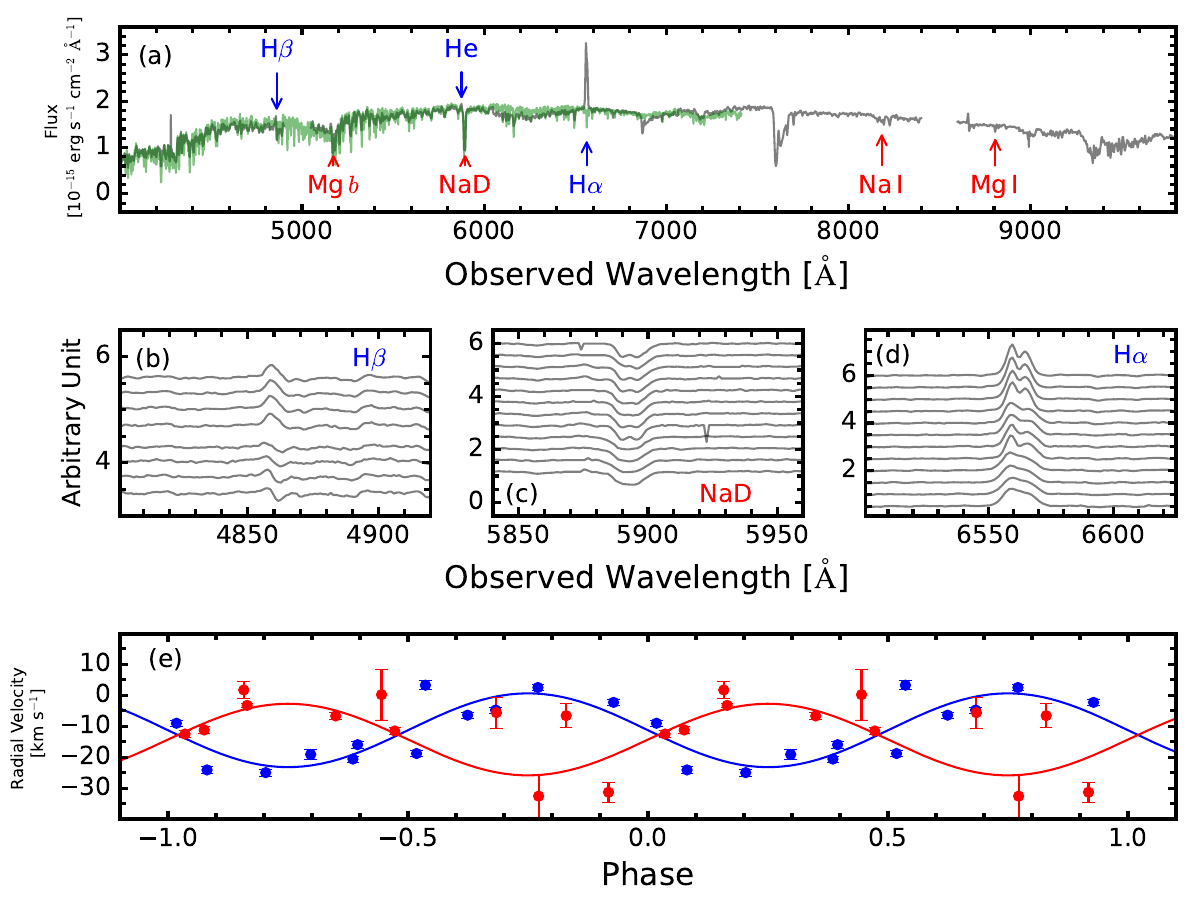}
\caption{(Top panel) Extinction-corrected and stacked 
GMOS-S spectrum (grey color)
of \target\ compared with a model spectrum (green color)
from the MILES library for a star 
with T$_{\rm eff} = 4570$\,K, ${\rm [Fe/H]} = 0.06$ dex, 
and $log\,g = 4.5$.
Notable emission and absorption features are marked  with arrows.
(Middle panels)
A sequence of the enlarged individual spectra centered on
H$\beta$ emission, {\nad} absorption, and H$\alpha$ emission features
from left to right, respectively, 
arranged in time sequence from top spectrum to bottom spectrum.
(Bottom panel) The radial velocity curves of \target\
obtained by H$\alpha$ emission (blue circles) and {\nad} absorption (red circles) compared with 
the best-fit sinusoidal curves shown in blue and red lines, 
respectively. 
}
\label{combspec}
\end{figure*}

Figure~\ref{combspec} shows the extinction-corrected integrated
GMOS-S spectrum of \target, revealing 
a continuum shape that appears consistent with 
that of a late-type star as well as a very prominent {\ha} line 
and notable absorption lines of {\mgb}, {\nad}, and {\nai}.
We note that \hei\ emission at 5876${\rm \AA}$ and \ha\ emission are weakly and strongly detected in this source, respectively.
The He/H flux ratio of \target\ is 0.27 which is smaller than known dwarf novae ($>$0.5) below the period minimum \citep{Ken15}.
In order to estimate stellar parameters of the secondary, we use the Medium-resolution Isaac Newton Telescope Library of Empirical Spectra (MILES), an empirical stellar  library constructed by 985 observed stars \citep{San06}.
The grids of the stellar parameters in the library are in the range of $2747 < {\rm T}_{\rm eff}\,[{\rm K}] < 36000$, $-0.2 < {\rm log}\,g < 5.5$, and $-2.86 < {\rm [Fe/H]\,[dex]} < 1.65$, and they are densely sampled around ${\rm [Fe/H]} \sim 0.0$ dex, ${\rm log}\,g \sim 4.0$, and T$_{\rm eff} \sim $3500--7000\,K with mean grid spacing of $\Delta$T$_{\rm eff} = 32\,{\rm K}$, $\Delta{\rm log}\,g=0.06$, and $\Delta{\rm [Fe/H]}=0.01$ dex.
We find the best-fit library model to the observed spectrum of \target\ in the wavelength range of 4000{\AA}--7400{\AA} to be that of T$_{\rm eff} = 4570 \pm 40\,{\rm K}$, ${\rm [Fe/H]} \sim 0.06 \pm 0.15\, {\rm dex}$, and ${\rm log}\,g = 4.5 \pm 0.1$ with a reduced chi-square value of 0.99.
We exclude the wavelength ranges for the emission lines and telluric lines in our fitting.

We measure the radial velocities of the observed \ha\ and \nad\ lines by conducting double Gaussian fits to the individual spectra (see Figure 3 (b)--(d)).
For the \ha\ line, we first determine the separation between the peak wavelengths of the two Gaussians to be 5.92{\AA} and 7.24{\AA} for the first 4 spectra obtained on January 16, 2018 and the remaining 8 spectra obtained on January 24, 2018, respectively.
In this process, all fitting parameters are set free.
We then fix the separation between two Gaussian components to these values (i.e., 5.92{\AA} for the first 4 spectra and 7.24{\AA} for the other 8 spectra) and conduct another double Gaussian fit to obtain the peak wavelengths of the two Gaussian components, as was done in previous studies of dwarf nova emission lines \citep[e.g.][]{Pat08,Gre20}.
For the \nad\ lines, we fix the wavelength separation of the two Gaussian components to be its intrinsic value of 5.97{\AA} to obtain the other fitting parameters of the peak wavelengths and intensities. A Long-Scargle analysis\footnote{astropy.stats.LombScargle is implemented \citep[\url{http://docs.astropy.org/en/stable/stats/lombscargle.html};][]{Ast13,Ast18}} of the fitted wavelengths of the \ha\ and \nad\ lines of the 12 spectra gives 58.78 minutes (or 24.5 cycles day$^{-1}$) as the binary orbital period of {\target}.

By adopting this as the initial value for the orbital period 
in the sinusoidal fitting of the \ha\ velocity evolution, we measure 
the orbital period of \target\ to be 58.75 $\pm$ 0.02 minutes 
as shown in the bottom panel of Figure~\ref{combspec}. 
Note that we obtain the same orbital period ($58.70\pm0.03$ minutes) when we conduct the same fitting to \nad\ velocities,
but with less statistical significance due to 
the increased uncertainties of their velocities.
The radial velocity amplitude derived for \target\ from
the observed \ha\ lines is $11.9\pm2.9${\kms},
which is relatively small but still belongs to the range previously
observed in other dwarf novae below the period minimum \citep{Aug96,Pat08,Bre12,Ken15,Gre20}.
The small amplitude is likely due to a small inclination angle of the source.

\section{Distance and Extinction} \label{sec:extinc}
\target\ has been observed by the GAIA satellite \citep{Gai16,Gai18}
to have a parallax of 0.714 $\pm$ 0.064 mas, 
providing its distance of $1.40 \pm 0.13$ kpc.

\begin{figure}
\epsscale{1.2}
\plotone{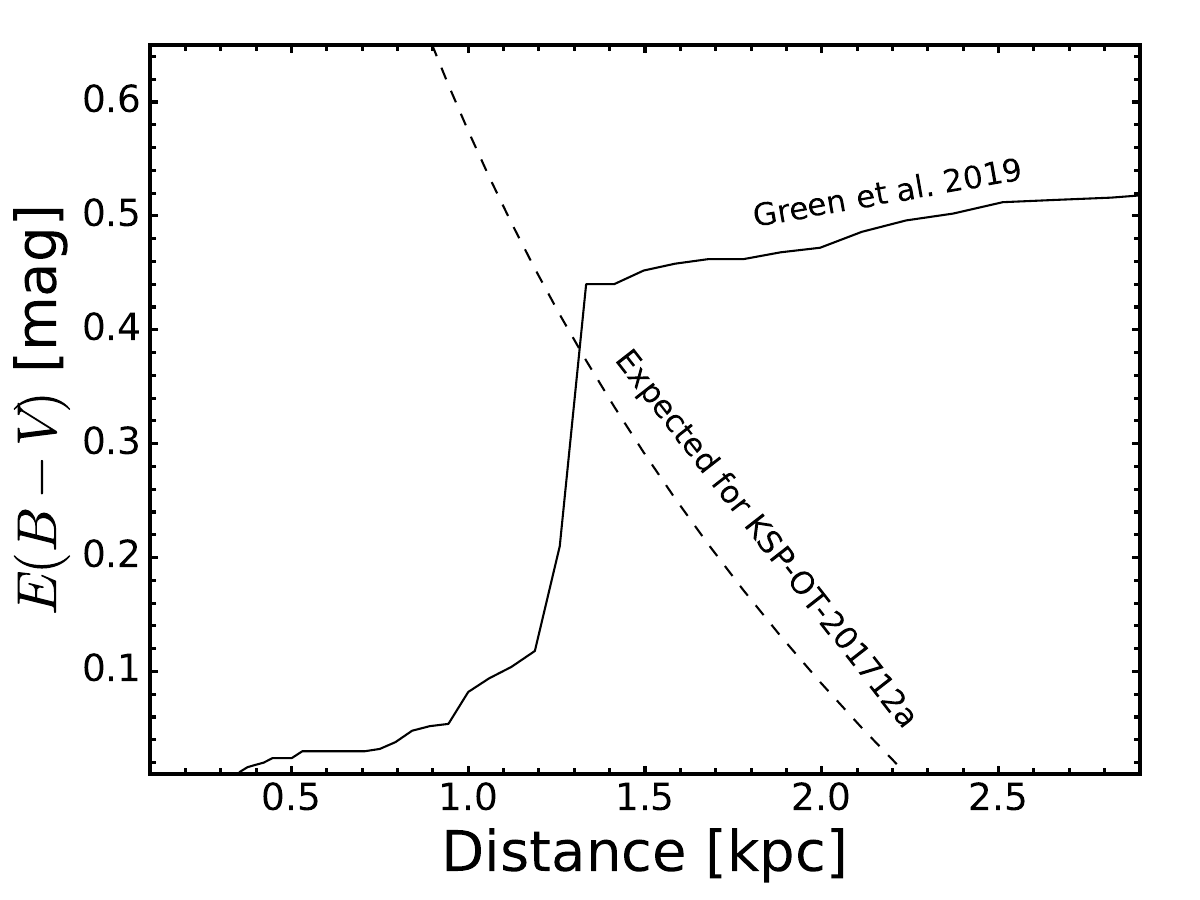}
\caption{The solid curve represents the distribution 
of the expected $E$(\bv) values as a function of 
distance in the direction of \target\ from the 3D extinction model of \citet{Gre19}.
The dashed curve represents the distribution of the
$E$(\bv) values that satisfy the known relation
between the orbital periods and the absolute peak
magnitudes of dwarf novae for \target.
}
\label{ebvdist}
\end{figure}

The peak luminosities of dwarf nova outbursts are known
to be correlated with their binary orbital periods  \citep[e.g.,][]{War95,Pat11}.
Using the orbital period of 58.75 minutes for \target\ (\S~\ref{sec:spec}), 
and we obtain 5.42 $\pm$ 0.47 mag as the peak $V$-band absolute magnitude 
of the source based on the relation between luminosities and periods in \citet[][see equation 3 for normal outbursts]{Pat11}.
The peak $V$-band apparent magnitude of the O8 outburst, which is a well-isolated outburst between quiescence (Figure~\ref{fulllc}), 
is 17.20 $\pm$ 0.02 mag, we attribute the difference of 11.78 mag to the distance and extinction to the source.
If we combine this and the 3D $E(B-V)$ model by \citet{Gre19} in the direction of {\target},
the best-fit combination of distance and extinction values are 
1.32 $\pm$ 0.10 kpc and $E(B-V)$ = 0.38 $\pm$ 0.10 mag (see Figure~\ref{ebvdist}).
The peak magnitude of the O7 outburst, which is $17.80 \pm 0.02$ mag, corresponds to the distance of $1.56 \pm 0.02$ kpc and E($B-V$) of $0.46 \pm 0.01$ mag.
These are slightly larger that those of the O8 outburst. Given that O8 outburst was observed with higher cadence around the peak than O7 outburst, we adopt the distance obtained using the O8 outburst in our study.
This distance estimation is very similar to that from the parallax measurement,
confirming that the isolated outburst O8 is likely a typical normal outburst of dwarf novae.
We adopt 1.32 $\pm$ 0.10 kpc as the distance to \target\ in this paper.
In addition, the extinction value $E(B-V)$ = 0.38 $\pm$ 0.10 mag 
is also in good agreement with the observed Balmer decrement of {\ha} and {\hb} from the source  (Figure~\ref{combspec})
where the ratio {\ha}/{\hb} $\simeq$ 4.68 translates into  $E(B-V)$ $\simeq$ 0.42 mag
by the model of \citet{Car89} in the case of a temperature $T=10^4$ K and an electron density $n_e=10^2$ cm$^{-3}$,
intrinsic Balmer decrement of {\ha}/{\hb} $=$ 2.86 \citep{Ost89}.
As the extinction to \target\ in this paper, we use $A_B = 1.57$, $A_V=1.18$, and $A_I=0.72$\, mag which are calculated with the $E(B-V)$ = $0.38\pm0.10$\,mag, $R_V = 3.1$ mag, and the extinction curve of \citet{Car89}.

\section{Discussion} \label{sec:discussion}

\subsection{The nature of {\target} and its outbursts}\label{sec:ori}

Based on the short period ($\simeq$59 minutes) of \target\ below the period gap, the presence of
standstills (see below) in its light curves caused by large mass-transfer rate, 
and the observed outburst cycles and amplitudes (Figure~\ref{fulllc}),
we classify the source to be an ER UMa-type dwarf nova as follows.
Standstills, which have been observed in ER UMa types for short-period dwarf novae,
are intermediate phases following an outburst peak with flat brightness 
that is about 0.7 mag fainter than the peak brightness \citep{Bua01,Szk13}.
Their duration ranges from as short as a few weeks 
to as large as more than a year \citep{Ohs23}. 
During the multiple outbursts O1--O6, the brightness of \target\ stalls 
at intermediate magnitudes of about two days
that are 0.8--1 mag fainter than the peak magnitudes 
after each outburst peak,
consistent with the behavior of short-interval standstills 
observed in other ER UMa type dwarf novae.
We, therefore, identify them to be standstills.
In terms of outburst pattern, the $\sim$6 and $\sim$15 days of the observed outburst cycles 
of O1--O6 and O7--O8, respectively, for \target\ 
and their amplitudes of $\sim$1.24 mag (O1--O6) and $\sim$1.60 mag (O7--O8), on the other hand, 
belong to the cycle ranges of 3--15 days and amplitudes of 1--4 mag
that have been observed in ER UMa types \citep{Otu16}.
Its orbital period of $\simeq$59 minutes is considerably smaller than
the observed range of 1.25--1.58 hours for ER UMa types \citep{Otu16}, 
and in fact, \target\ is an ER UMa dwarf nova with the shortest 
orbital period as far as we are aware.

Standstills, which have been observed in ER UMa types 
and its long-period version of Z Cam types, 
are known to be caused by a large mass-transfer rate \citep{Bua01,Ohs23}.
However, a dwarf nova with a cluster of recurring standstills 
(as seen in the O1--O6 period of \target),
maintaining the minimum brightness substantially above the quiescent level,
is extremely rare since dwarf novae return to a quiescent phase 
after a standstill \citep{Ohs23}. 
Even in the case of the few dwarf novae 
that show a standstill followed by an outburst activity
instead of immediately returning to a quiescent phase \citep{Sim11,Szk13,Kat19a,Kat19c},
the outburst activity is very simple and composed of a single outburst.
\target\ is, therefore, a unique dwarf nova showing the presence of
a cluster of standstills, i.e., O1--O6, above the quiescent level.

The origin of standstills is relatively unknown, 
although attempts have been made to attribute it to 
changes in mass-transfer rate from the secondary  \citep[e.g.,][]{Ham14} 
or to limit-cycle oscillations between 
standstills from the inner-disk area and outbursts 
caused by thermal instability from the outer-disk \citep[e.g.,][]{Kat19c}.
What initiates the required changes in the mass-transfer rate and
the limit-cycle oscillations, however, remains uncertain.
We note that the relatively periodic ($\sim$6.6 days) 
burst activities of O1--O6 of \target\ can be more readily
explainable by the limit-cycle oscillation process
in the frame of thermal disk instability \citep{Can87}.

It is known that there exists a good correlation between 
the average $V$-band magnitudes and the mass accretion rates
of dwarf novae \citep[e.g.,][]{Pac80,Dub18}.
Applying the correlation to the observed brightness of the outbursts of {\target}, we obtain the expected accretion rates in the range of $\sim$0.4--2.6: $0.9$ (O1), $1.9$ (O2), $2.6$ (O3), $1.6$ (O4), $1.4$ (O5), $0.6$ (O6), $0.4$ (O7) and $0.5$ (O8), all in unit of $10^{17}$ g\, s$^{-1}$.
The inferred critical mass is $\sim0.35\times10^{17}$ g\, s$^{-1}$ using its orbital period of 59 minutes and the relation $\dot{M}_{\rm crit}  \sim 3.5\times10^{16}P_{\rm orb}^{1.6}$ g\,s$^{-1}$ from \citet{Dub18}.
This value is similar to the expected accretion rates of the four outbursts of O1, O6, O7 and O8, while it is somewhat smaller than those of the other four outburst of O2, O3, O4 and O5.
This is consistent with the presence of standstills in the latter four outbursts.

In Figure~\ref{zoomlc}, the \bv\ and \vi\ colors of \target\ show different 
evolutions between O2--O5 and O7--O8 outbursts in which the former 
become bluest at their brightness peaks, 
as most of the other outbursts observed in dwarf novae \citep[][e.g.,]{Pic21},
while the latter do the opposite pattern. 
O1, which precedes O2--O5 outbursts, is similar to O7--O8 in color.
The colors of an accretion disk during outbursts are largely determined by the 
internal heat wave transfer and the accretion rate of cold material from outside.
A small accretion rate creates outbursts at the inner part of the accretion
and tends to be followed by a slow internal heat wave transfer 
to the outside of the accretion disk \citep{Can86}.
The observed colors of O7--O8, 
as well as those of O1 and O6 which are closer to O7--O8 than O2--O5 (\S\ref{sec:discovery}),
suggest that the internal heat wave transfer rate
is less efficient than the accretion rate of cold material,
leading to a redward evolution, consistent with their small accretion rate.
In the case of the O2--O5 outbursts, 
their large accretion rate is more compatible with outbursts at the
outer part of the accretion disk and an efficient outside-in heat wave transfer,
leading to a blueward evolution if the heat wave transfer effect is more dominant
than the accretion rate of cold material in their colors.

\subsection{Evolution of {\target}} \label{sec:zcam}
\begin{figure}
\epsscale{1.2}
\plotone{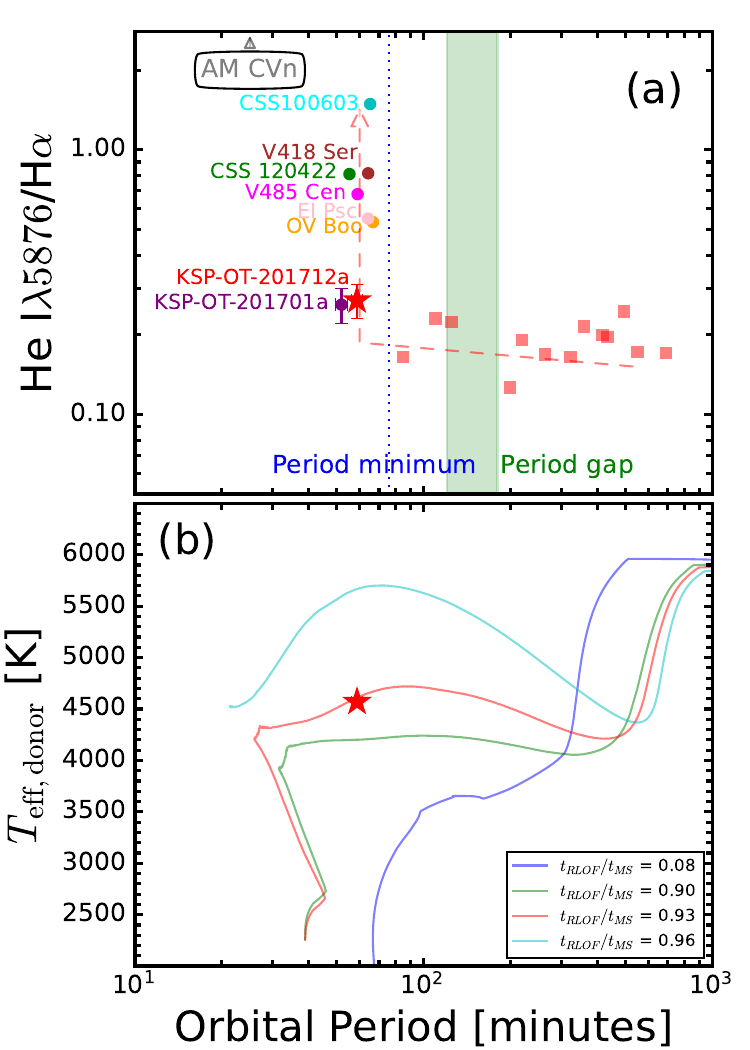}
\caption{
(a) Observed orbital periods and He/H ratios of dwarf novae: red squares for long-period dwarf novae; circles for short-period dwarf novae; and red star for {\target}.
The green shared area shows the period gap.
The blue dotted vertical line indicates the period minimum of 76 minutes.
The arrowed red dashed curve roughly represents the expected evolutionary path of a dwarf nova with an evolved secondary as its orbital period decreases.
The AM CVn systems, which are a group of dwarf novae with small orbital periods, small mass ratios, and no H emission, are located at the top-left corner. 
The orbital periods and He/H ratios of other dwarf novae are from \citet{Lee22}.
(b) Calculated effective temperatures of the secondary star of \target\ by MESA (see text) as a function of orbital period: the blue curve is for the case in which the initial mass transfer started when its age was 8\% of the main sequence.
The green, red, and cyan curves are the same as the blue curve but for the cases of 90, 93, and 96\%, respectively.
The red star marks the position of {\target}.
}
\label{hetemp}
\end{figure}

Short-period dwarf novae below the period minimum with a low He/H flux ratio
are of important consequence to our understanding of how low-mass binaries
evolve to He CVs or {\amcvn}, the two ultimate phases of dwarf novae 
prior to Type Ia SN explosions, 
from the long-period dwarf nova phase \citep{Bil07}. 
Note that only a handful of dwarf novae have been observed 
below the period minimum ($\sim$75 minutes), and 
most of them have a large He/H flux ratio ($>$0.5)
due to the small amount of hydrogen in their secondary stars \citep[e.g.,][]{Ken15}.
Little has been known about short-period dwarf novae that still have
a substantial amount of hydrogen (= a low He/H flux ratio)
from an early evolutionary phase of low-mass binaries
below the period minimum. 
As we reported previously in \citet{Lee22},
until now, \targetold\ has been the only short-period dwarf nova  
under the period minimum with a small He/H flux ratio of 0.26.
Although the discovery of \targetold\ did confirm the presence 
of a transitional phase between long-period dwarf novae
and He CVs in the evolutionary track of low-mass binaries,
its spectrum was dominated by the emission from the accretion disk.
This made it impossible to conduct further investigation into 
the physical conditions of the source such as the size and
temperature of its secondary star
which is expected to play a pivotal
role in their evolution \citep[e.g.,][]{Tho02a}.
{\target}, with an orbital period of 58.75 minutes 
and He/H flux ratio of $\sim$0.27,
is therefore the second dwarf nova with the transitional nature
of a short period and low He/H flux ratio and the first one 
with a measured surface temperature of the secondary, 
providing a unique opportunity to study 
its density condition and mass transfer history 
during this intermediate phase
in the evolution of low-mass binaries \citep{Pod03}.
Figure~\ref{hetemp} (a) compares the orbital period and He/H ratio of \target\ with those of other dwarf novae, showing that the source and KSP-OT-201701a together correspond to the transitional phase between long- and short-period dwarf.

It is known that dwarf novae with higher effective temperatures 
at a given orbital period have evolved donor stars \citep{Tho02a}.
The measured $\sim$4570~K temperature of the secondary star
of \target\ is greater than the 3000--3500~K range in which 
dwarf novae whose orbital periods are slightly greater 
than the period minimum have been measured \citep{Beu98,Tho02a},
while it is lower than $\sim$6700~K for He CVs \citep{Bur22}.
In the case of He CVs, which is a more evolved short-period
dwarf nova with a larger He/H flux ratio than \target,
the measured temperatures of $\sim$6700~K for the secondaries
are in good agreement with the theoretical prediction
made by stellar evolution code MESA \citep[Modules for Experiments in Stellar Astrophysics;][]{Pax11,Pax13,Pax15,Pax18,Pax19}.
No such comparison between theoretical predictions and observational
examples has been made 
for short-period dwarf novae with a low He/H flux ratio so far
to the best of our knowledge.

Figure~\ref{hetemp} (b) shows the evolutionary tracks of the effective temperature of the secondary of \target\ that we calculate using MESA \citep[version r22.11.1; following a description of][]{ElB21a} for four different cases of the initial mass transfer age. As shown in the figure, we find that the mass transfer in \target\ started when its age was about 93\% of the main sequence age.
Theoretically, the secondary stars of dwarf novae 
passing through the period minimum are generally expected to start mass 
transfer when their ages are over $\sim$80\% of the main sequence age \citep{Gol15,ElB21a,Bur22}.
We, therefore, find that the temperature of the secondary star
of \target\ is within the range that is compatible with 
the theoretical predictions and that its age was in fact about 
13\% older than the minimum main-sequence age for mass transfer to start
and pass through the period minimum.
Following our binary evolutionary model 
that corresponds to \target, 
the source is expected to evolve to 
the \amcvn\ phase with an orbital period of as short as 20 minutes
after which its period bounces back a bit
when the thermal time scale of the secondary star becomes larger than gravitational radiation time scale \citep{Rap82}.

\target\ is a unique example showing that standstills can 
persist in a dwarf nova to the dynamical evolutionary stage 
after the passage of the period minimum
with atypical outbursts embedded in standstills.
To the best of our knowledge, no dwarf novae below the period
minimum have been observed with standstills.
Also, standstills in general appear as a middle stage
from an outburst to the quiescent phase, which is different
from the standstill in \target. 
The secondary of \target\ still has a significant hydrogen envelope
revealed by the strong \ha\ emission (Figure~\ref{combspec}), 
and the mass-accretion rate of the system is still largely
supported by the presence of standstills and short outburst cycles (Figure~\ref{fulllc}).
The secondary will gradually exhaust its hydrogen envelope 
via mass transfer and the system will enter a dwarf nova phase
that has outburst cycles without standstills \citep{Lee22}.
Only a handful of dwarf novae have been observed 
below the period minimum, and more samples are required
to understand the evolution of short-period dwarf novae.

\section{Summary and Conclusion} \label{sec:summary}
\target\ is the first He-deficient dwarf nova 
below the period minimum with a measured
temperature for a secondary. 
We summarize the results as follows.

\begin{itemize}

\item The source exhibits peculiar light curves
for a dwarf nova with outbursts embedded in standstills,
which identifies it to be an ER-UMa type undergoing substantial mass transfer.
The outburst cycle during the standstills is 
$\sim$ 6.6 days but it increases to 
$\sim$ 15 days outside of them.

\item The outburst colors become bluest near the
peaks during the standstills, but they evolve
in the opposite direction outside of the standstills.
We attribute this to the difference in the mass 
transfer rate, in which outside-in outburst processes
with an increased mass transfer are responsible for the
outburst within the standstills while inside-out
outburst processes with a smaller mass transfer are
responsible for those outside of the standstills.
    
\item Using radial velocity curves obtained
from time series spectroscopy,
we estimate its orbital period to be
$58.75 \pm 0.02$ minutes, which is below
the period minimum. 
The spectral energy distribution of the source
appears to be very similar to that of a late-type star, apparently dominated by its secondary.
While its double-peaked H$\alpha$ emission is very strong,
He emission is weak, giving the He/H flux ratio
of $\sim$ 0.27 and making the source 
a He-deficient dwarf nova.
By analyzing the absorption lines of {\nad}, {\mgb}, and {\nai}, 
we obtain the following parameters for its secondary: 
T$_{\rm eff} \simeq 4570 \pm 40$ K, ${\rm [Fe/H]} \simeq 0.06 \pm 0.15$ dex and log~$g \simeq 4.5 \pm 0.1$.
Its orbital period and outburst peak, as well as 
its measured proper motion, are consistent
with the source being at the distance of  $1.32\pm0.10$ kpc. 
	
\item Long-period dwarf novae above the period minimum typically show 
strong {\ha} but weak He emission, 
while short-period dwarf novae below the period minimum 
have weak {\ha} emission but strong He emission.
{\target} is a He-deficient (or, small He/H flux ratio), short-period dwarf nova below the period minimum
showing a transitional nature evolving from 
a long-period dwarf nova to a short-period one. 
The hot secondary suggests that the mass transfer in \target\ started when 
its age was about 93\% of the main sequence age with a large central density. 
It is expected that \target\ will eventually evolve into an \amcvn.
	
\end{itemize}

\begin{acknowledgments}
This research has made use of the KMTNet system operated by the Korea Astronomy and Space Science Institute (KASI) 
at three host sites of CTIO in Chile, SAAO in South Africa, and SSO in Australia.
Data transfer from the host site to KASI was supported by the Korea Research Environment Open NETwork (KREONET).
The Gemini South observations were obtained under the K-GMT Science Program (PID: GS-2017B-Q-21) of KASI.
This research was supported by the Korea Astronomy and Space Science Institute
under the R\&D program (Project No. 2024-1-860-02) supervised by the Ministry of Science and ICT.
Y. Lee was supported by Basic Science Research Program through the National Research Foundation of Korea (NRF) funded by the Ministry of Education (NRF-2022R1I1A1A01054555). 
D.S.M was supported in part by a Leading Edge Fund from the Canadian Foundation for Innovation (project No. 30951) and a Discovery Grant (RGPIN2019-06524) from the Natural Sciences and Engineering Research Council (NSERC) of Canada.

\end{acknowledgments}

\bibliography{draft}

\begin{thebibliography}{}
\expandafter\ifx\csname natexlab\endcsname\relax\def\natexlab#1{#1}\fi
\providecommand{\url}[1]{\href{#1}{#1}}
\providecommand{\dodoi}[1]{doi:~\href{http://doi.org/#1}{\nolinkurl{#1}}}
\providecommand{\doeprint}[1]{\href{http://ascl.net/#1}{\nolinkurl{http://ascl.net/#1}}}
\providecommand{\doarXiv}[1]{\href{https://arxiv.org/abs/#1}{\nolinkurl{https://arxiv.org/abs/#1}}}

\bibitem[{{Afsariardchi} {et~al.}(2019){Afsariardchi}, {Moon}, {Drout},
  {Gonz{\'a}lez-Gait{\'a}n}, {Ni}, {Matzner}, {Kim}, {Lee}, {Park}, {Gal-Yam},
  {Pignata}, {Koo}, {Ryder}, {Cha}, \& {Lee}}]{Afs19}
{Afsariardchi}, N., {Moon}, D.-S., {Drout}, M.~R., {et~al.} 2019, \apj, 881,
  22, \dodoi{10.3847/1538-4357/ab2be6}

\bibitem[{{Antoniadis} {et~al.}(2017){Antoniadis}, {Moon}, {Ni}, {Kim}, {Lee},
  \& {Neilson}}]{Ant17}
{Antoniadis}, J., {Moon}, D.-S., {Ni}, Y.~Q., {et~al.} 2017, \apj, 844, 160,
  \dodoi{10.3847/1538-4357/aa706b}

\bibitem[{{Astropy Collaboration} {et~al.}(2013){Astropy Collaboration},
  {Robitaille}, {Tollerud}, {Greenfield}, {Droettboom}, {Bray}, {Aldcroft},
  {Davis}, {Ginsburg}, {Price-Whelan}, {Kerzendorf}, {Conley}, {Crighton},
  {Barbary}, {Muna}, {Ferguson}, {Grollier}, {Parikh}, {Nair}, {Unther},
  {Deil}, {Woillez}, {Conseil}, {Kramer}, {Turner}, {Singer}, {Fox}, {Weaver},
  {Zabalza}, {Edwards}, {Azalee Bostroem}, {Burke}, {Casey}, {Crawford},
  {Dencheva}, {Ely}, {Jenness}, {Labrie}, {Lim}, {Pierfederici}, {Pontzen},
  {Ptak}, {Refsdal}, {Servillat}, \& {Streicher}}]{Ast13}
{Astropy Collaboration}, {Robitaille}, T.~P., {Tollerud}, E.~J., {et~al.} 2013,
  \aap, 558, A33, \dodoi{10.1051/0004-6361/201322068}

\bibitem[{{Astropy Collaboration} {et~al.}(2018){Astropy Collaboration},
  {Price-Whelan}, {Sip{\H o}cz}, {G{\"u}nther}, {Lim}, {Crawford}, {Conseil},
  {Shupe}, {Craig}, {Dencheva}, {Ginsburg}, {VanderPlas}, {Bradley},
  {P{\'e}rez-Su{\'a}rez}, {de Val-Borro}, {Aldcroft}, {Cruz}, {Robitaille},
  {Tollerud}, {Ardelean}, {Babej}, {Bach}, {Bachetti}, {Bakanov}, {Bamford},
  {Barentsen}, {Barmby}, {Baumbach}, {Berry}, {Biscani}, {Boquien}, {Bostroem},
  {Bouma}, {Brammer}, {Bray}, {Breytenbach}, {Buddelmeijer}, {Burke},
  {Calderone}, {Cano Rodr{\'{\i}}guez}, {Cara}, {Cardoso}, {Cheedella},
  {Copin}, {Corrales}, {Crichton}, {D'Avella}, {Deil}, {Depagne}, {Dietrich},
  {Donath}, {Droettboom}, {Earl}, {Erben}, {Fabbro}, {Ferreira}, {Finethy},
  {Fox}, {Garrison}, {Gibbons}, {Goldstein}, {Gommers}, {Greco}, {Greenfield},
  {Groener}, {Grollier}, {Hagen}, {Hirst}, {Homeier}, {Horton}, {Hosseinzadeh},
  {Hu}, {Hunkeler}, {Ivezi{\'c}}, {Jain}, {Jenness}, {Kanarek}, {Kendrew},
  {Kern}, {Kerzendorf}, {Khvalko}, {King}, {Kirkby}, {Kulkarni}, {Kumar},
  {Lee}, {Lenz}, {Littlefair}, {Ma}, {Macleod}, {Mastropietro}, {McCully},
  {Montagnac}, {Morris}, {Mueller}, {Mumford}, {Muna}, {Murphy}, {Nelson},
  {Nguyen}, {Ninan}, {N{\"o}the}, {Ogaz}, {Oh}, {Parejko}, {Parley}, {Pascual},
  {Patil}, {Patil}, {Plunkett}, {Prochaska}, {Rastogi}, {Reddy Janga},
  {Sabater}, {Sakurikar}, {Seifert}, {Sherbert}, {Sherwood-Taylor}, {Shih},
  {Sick}, {Silbiger}, {Singanamalla}, {Singer}, {Sladen}, {Sooley},
  {Sornarajah}, {Streicher}, {Teuben}, {Thomas}, {Tremblay}, {Turner},
  {Terr{\'o}n}, {van Kerkwijk}, {de la Vega}, {Watkins}, {Weaver}, {Whitmore},
  {Woillez}, {Zabalza}, \& {Astropy Contributors}}]{Ast18}
{Astropy Collaboration}, {Price-Whelan}, A.~M., {Sip{\H o}cz}, B.~M., {et~al.}
  2018, \aj, 156, 123, \dodoi{10.3847/1538-3881/aabc4f}

\bibitem[{{Augusteijn} {et~al.}(1996){Augusteijn}, {van der Hooft}, {de Jong},
  \& {van Paradijs}}]{Aug96}
{Augusteijn}, T., {van der Hooft}, F., {de Jong}, J.~A., \& {van Paradijs}, J.
  1996, \aap, 311, 889

\bibitem[{{Bellm} {et~al.}(2019){Bellm}, {Kulkarni}, {Graham}, {Dekany},
  {Smith}, {Riddle}, {Masci}, {Helou}, {Prince}, {Adams}, {Barbarino},
  {Barlow}, {Bauer}, {Beck}, {Belicki}, {Biswas}, {Blagorodnova}, {Bodewits},
  {Bolin}, {Brinnel}, {Brooke}, {Bue}, {Bulla}, {Burruss}, {Cenko}, {Chang},
  {Connolly}, {Coughlin}, {Cromer}, {Cunningham}, {De}, {Delacroix}, {Desai},
  {Duev}, {Eadie}, {Farnham}, {Feeney}, {Feindt}, {Flynn}, {Franckowiak},
  {Frederick}, {Fremling}, {Gal-Yam}, {Gezari}, {Giomi}, {Goldstein},
  {Golkhou}, {Goobar}, {Groom}, {Hacopians}, {Hale}, {Henning}, {Ho}, {Hover},
  {Howell}, {Hung}, {Huppenkothen}, {Imel}, {Ip}, {Ivezi{\'c}}, {Jackson},
  {Jones}, {Juric}, {Kasliwal}, {Kaspi}, {Kaye}, {Kelley}, {Kowalski},
  {Kramer}, {Kupfer}, {Landry}, {Laher}, {Lee}, {Lin}, {Lin}, {Lunnan},
  {Giomi}, {Mahabal}, {Mao}, {Miller}, {Monkewitz}, {Murphy}, {Ngeow},
  {Nordin}, {Nugent}, {Ofek}, {Patterson}, {Penprase}, {Porter}, {Rauch},
  {Rebbapragada}, {Reiley}, {Rigault}, {Rodriguez}, {van Roestel}, {Rusholme},
  {van Santen}, {Schulze}, {Shupe}, {Singer}, {Soumagnac}, {Stein}, {Surace},
  {Sollerman}, {Szkody}, {Taddia}, {Terek}, {Van Sistine}, {van Velzen},
  {Vestrand}, {Walters}, {Ward}, {Ye}, {Yu}, {Yan}, \& {Zolkower}}]{Bel19}
{Bellm}, E.~C., {Kulkarni}, S.~R., {Graham}, M.~J., {et~al.} 2019, \pasp, 131,
  018002, \dodoi{10.1088/1538-3873/aaecbe}

\bibitem[{{Belloni} \& {Schreiber}(2023)}]{Bel23}
{Belloni}, D., \& {Schreiber}, M.~R. 2023, \aap, 678, A34,
  \dodoi{10.1051/0004-6361/202347047}

\bibitem[{{Beuermann} {et~al.}(1998){Beuermann}, {Baraffe}, {Kolb}, \&
  {Weichhold}}]{Beu98}
{Beuermann}, K., {Baraffe}, I., {Kolb}, U., \& {Weichhold}, M. 1998, \aap, 339,
  518

\bibitem[{{Bildsten} {et~al.}(2007){Bildsten}, {Shen}, {Weinberg}, \&
  {Nelemans}}]{Bil07}
{Bildsten}, L., {Shen}, K.~J., {Weinberg}, N.~N., \& {Nelemans}, G. 2007,
  \apjl, 662, L95, \dodoi{10.1086/519489}

\bibitem[{{Breedt} {et~al.}(2012){Breedt}, {G{\"a}nsicke}, {Marsh}, {Steeghs},
  {Drake}, \& {Copperwheat}}]{Bre12}
{Breedt}, E., {G{\"a}nsicke}, B.~T., {Marsh}, T.~R., {et~al.} 2012, \mnras,
  425, 2548, \dodoi{10.1111/j.1365-2966.2012.21724.x}

\bibitem[{{Brown} {et~al.}(2018){Brown}, {Moon}, {Ni}, {Drout}, {Antoniadis},
  {Afsariardchi}, {Cha}, \& {Lee}}]{Bro18}
{Brown}, S., {Moon}, D.-S., {Ni}, Y.~Q., {et~al.} 2018, \apj, 860, 21,
  \dodoi{10.3847/1538-4357/aabfe2}

\bibitem[{{Buat-M{\'e}nard} {et~al.}(2001{\natexlab{a}}){Buat-M{\'e}nard},
  {Hameury}, \& {Lasota}}]{Bua01b}
{Buat-M{\'e}nard}, V., {Hameury}, J.~M., \& {Lasota}, J.~P. 2001{\natexlab{a}},
  \aap, 369, 925, \dodoi{10.1051/0004-6361:20010176}

\bibitem[{{Buat-M{\'e}nard} {et~al.}(2001{\natexlab{b}}){Buat-M{\'e}nard},
  {Hameury}, \& {Lasota}}]{Bua01}
{Buat-M{\'e}nard}, V., {Hameury}, J.-M., \& {Lasota}, J.-P. 2001{\natexlab{b}},
  \aap, 366, 612, \dodoi{10.1051/0004-6361:20000107}

\bibitem[{{Burdge} {et~al.}(2022){Burdge}, {El-Badry}, {Marsh}, {Rappaport},
  {Brown}, {Caiazzo}, {Chakrabarty}, {Dhillon}, {Fuller}, {G{\"a}nsicke},
  {Graham}, {Kara}, {Kulkarni}, {Littlefair}, {Mr{\'o}z}, {Rodr{\'\i}guez-Gil},
  {Roestel}, {Simcoe}, {Bellm}, {Drake}, {Dekany}, {Groom}, {Laher}, {Masci},
  {Riddle}, {Smith}, \& {Prince}}]{Bur22}
{Burdge}, K.~B., {El-Badry}, K., {Marsh}, T.~R., {et~al.} 2022, \nat, 610, 467,
  \dodoi{10.1038/s41586-022-05195-x}

\bibitem[{{Cannizzo} \& {Kenyon}(1987)}]{Can87}
{Cannizzo}, J.~K., \& {Kenyon}, S.~J. 1987, \apj, 320, 319,
  \dodoi{10.1086/165545}

\bibitem[{{Cannizzo} {et~al.}(1986){Cannizzo}, {Wheeler}, \& {Polidan}}]{Can86}
{Cannizzo}, J.~K., {Wheeler}, J.~C., \& {Polidan}, R.~S. 1986, \apj, 301, 634,
  \dodoi{10.1086/163929}

\bibitem[{{Cardelli} {et~al.}(1989){Cardelli}, {Clayton}, \& {Mathis}}]{Car89}
{Cardelli}, J.~A., {Clayton}, G.~C., \& {Mathis}, J.~S. 1989, \apj, 345, 245,
  \dodoi{10.1086/167900}

\bibitem[{{Carter} {et~al.}(2013){Carter}, {Steeghs}, {de Miguel}, {Goff},
  {Koff}, {Krajci}, {Marsh}, {G{\"a}nsicke}, {Breedt}, {Groot}, {Nelemans},
  {Roelofs}, {Rau}, {Koester}, \& {Kupfer}}]{Car13}
{Carter}, P.~J., {Steeghs}, D., {de Miguel}, E., {et~al.} 2013, \mnras, 431,
  372, \dodoi{10.1093/mnras/stt169}

\bibitem[{{Dubus} {et~al.}(2018){Dubus}, {Otulakowska-Hypka}, \&
  {Lasota}}]{Dub18}
{Dubus}, G., {Otulakowska-Hypka}, M., \& {Lasota}, J.-P. 2018, \aap, 617, A26,
  \dodoi{10.1051/0004-6361/201833372}

\bibitem[{{El-Badry} {et~al.}(2021{\natexlab{a}}){El-Badry}, {Rix}, {Quataert},
  {Kupfer}, \& {Shen}}]{ElB21}
{El-Badry}, K., {Rix}, H.-W., {Quataert}, E., {Kupfer}, T., \& {Shen}, K.~J.
  2021{\natexlab{a}}, \mnras, 508, 4106, \dodoi{10.1093/mnras/stab2583}

\bibitem[{{El-Badry} {et~al.}(2021{\natexlab{b}}){El-Badry}, {Quataert}, {Rix},
  {Weisz}, {Kupfer}, {Shen}, {Xiang}, {Yang}, \& {Liu}}]{ElB21a}
{El-Badry}, K., {Quataert}, E., {Rix}, H.-W., {et~al.} 2021{\natexlab{b}},
  \mnras, 505, 2051, \dodoi{10.1093/mnras/stab1318}

\bibitem[{{Gaia Collaboration} {et~al.}(2016){Gaia Collaboration}, {Prusti},
  {de Bruijne}, {Brown}, {Vallenari}, {Babusiaux}, {Bailer-Jones}, {Bastian},
  {Biermann}, {Evans}, {Eyer}, {Jansen}, {Jordi}, {Klioner}, {Lammers},
  {Lindegren}, {Luri}, {Mignard}, {Milligan}, {Panem}, {Poinsignon},
  {Pourbaix}, {Randich}, {Sarri}, {Sartoretti}, {Siddiqui}, {Soubiran},
  {Valette}, {van Leeuwen}, {Walton}, {Aerts}, {Arenou}, {Cropper}, {Drimmel},
  {H{\o}g}, {Katz}, {Lattanzi}, {O'Mullane}, {Grebel}, {Holland}, {Huc},
  {Passot}, {Bramante}, {Cacciari}, {Casta{\~n}eda}, {Chaoul}, {Cheek}, {De
  Angeli}, {Fabricius}, {Guerra}, {Hern{\'a}ndez}, {Jean-Antoine-Piccolo},
  {Masana}, {Messineo}, {Mowlavi}, {Nienartowicz}, {Ord{\'o}{\~n}ez-Blanco},
  {Panuzzo}, {Portell}, {Richards}, {Riello}, {Seabroke}, {Tanga},
  {Th{\'e}venin}, {Torra}, {Els}, {Gracia-Abril}, {Comoretto},
  {Garcia-Reinaldos}, {Lock}, {Mercier}, {Altmann}, {Andrae}, {Astraatmadja},
  {Bellas-Velidis}, {Benson}, {Berthier}, {Blomme}, {Busso}, {Carry},
  {Cellino}, {Clementini}, {Cowell}, {Creevey}, {Cuypers}, {Davidson}, {De
  Ridder}, {de Torres}, {Delchambre}, {Dell'Oro}, {Ducourant}, {Fr{\'e}mat},
  {Garc{\'\i}a-Torres}, {Gosset}, {Halbwachs}, {Hambly}, {Harrison}, {Hauser},
  {Hestroffer}, {Hodgkin}, {Huckle}, {Hutton}, {Jasniewicz}, {Jordan},
  {Kontizas}, {Korn}, {Lanzafame}, {Manteiga}, {Moitinho}, {Muinonen},
  {Osinde}, {Pancino}, {Pauwels}, {Petit}, {Recio-Blanco}, {Robin}, {Sarro},
  {Siopis}, {Smith}, {Smith}, {Sozzetti}, {Thuillot}, {van Reeven}, {Viala},
  {Abbas}, {Abreu Aramburu}, {Accart}, {Aguado}, {Allan}, {Allasia},
  {Altavilla}, {{\'A}lvarez}, {Alves}, {Anderson}, {Andrei}, {Anglada Varela},
  {Antiche}, {Antoja}, {Ant{\'o}n}, {Arcay}, {Atzei}, {Ayache}, {Bach},
  {Baker}, {Balaguer-N{\'u}{\~n}ez}, {Barache}, {Barata}, {Barbier}, {Barblan},
  {Baroni}, {Barrado y Navascu{\'e}s}, {Barros}, {Barstow}, {Becciani},
  {Bellazzini}, {Bellei}, {Bello Garc{\'\i}a}, {Belokurov}, {Bendjoya},
  {Berihuete}, {Bianchi}, {Bienaym{\'e}}, {Billebaud}, {Blagorodnova},
  {Blanco-Cuaresma}, {Boch}, {Bombrun}, {Borrachero}, {Bouquillon}, {Bourda},
  {Bouy}, {Bragaglia}, {Breddels}, {Brouillet}, {Br{\"u}semeister},
  {Bucciarelli}, {Budnik}, {Burgess}, {Burgon}, {Burlacu}, {Busonero}, {Buzzi},
  {Caffau}, {Cambras}, {Campbell}, {Cancelliere}, {Cantat-Gaudin}, {Carlucci},
  {Carrasco}, {Castellani}, {Charlot}, {Charnas}, {Charvet}, {Chassat},
  {Chiavassa}, {Clotet}, {Cocozza}, {Collins}, {Collins}, {Costigan}, {Crifo},
  {Cross}, {Crosta}, {Crowley}, {Dafonte}, {Damerdji}, {Dapergolas}, {David},
  {David}, {De Cat}, {de Felice}, {de Laverny}, {De Luise}, {De March}, {de
  Martino}, {de Souza}, {Debosscher}, {del Pozo}, {Delbo}, {Delgado},
  {Delgado}, {di Marco}, {Di Matteo}, {Diakite}, {Distefano}, {Dolding}, {Dos
  Anjos}, {Drazinos}, {Dur{\'a}n}, {Dzigan}, {Ecale}, {Edvardsson}, {Enke},
  {Erdmann}, {Escolar}, {Espina}, {Evans}, {Eynard Bontemps}, {Fabre},
  {Fabrizio}, {Faigler}, {Falc{\~a}o}, {Farr{\`a}s Casas}, {Faye}, {Federici},
  {Fedorets}, {Fern{\'a}ndez-Hern{\'a}ndez}, {Fernique}, {Fienga}, {Figueras},
  {Filippi}, {Findeisen}, {Fonti}, {Fouesneau}, {Fraile}, {Fraser}, {Fuchs},
  {Furnell}, {Gai}, {Galleti}, {Galluccio}, {Garabato}, {Garc{\'\i}a-Sedano},
  {Gar{\'e}}, {Garofalo}, {Garralda}, {Gavras}, {Gerssen}, {Geyer}, {Gilmore},
  {Girona}, {Giuffrida}, {Gomes}, {Gonz{\'a}lez-Marcos},
  {Gonz{\'a}lez-N{\'u}{\~n}ez}, {Gonz{\'a}lez-Vidal}, {Granvik}, {Guerrier},
  {Guillout}, {Guiraud}, {G{\'u}rpide}, {Guti{\'e}rrez-S{\'a}nchez}, {Guy},
  {Haigron}, {Hatzidimitriou}, {Haywood}, {Heiter}, {Helmi}, {Hobbs},
  {Hofmann}, {Holl}, {Holland}, {Hunt}, {Hypki}, {Icardi}, {Irwin}, {Jevardat
  de Fombelle}, {Jofr{\'e}}, {Jonker}, {Jorissen}, {Julbe}, {Karampelas},
  {Kochoska}, {Kohley}, {Kolenberg}, {Kontizas}, {Koposov}, {Kordopatis},
  {Koubsky}, {Kowalczyk}, {Krone-Martins}, {Kudryashova}, {Kull}, {Bachchan},
  {Lacoste-Seris}, {Lanza}, {Lavigne}, {Le Poncin-Lafitte}, {Lebreton},
  {Lebzelter}, {Leccia}, {Leclerc}, {Lecoeur-Taibi}, {Lemaitre}, {Lenhardt},
  {Leroux}, {Liao}, {Licata}, {Lindstr{\o}m}, {Lister}, {Livanou}, {Lobel},
  {L{\"o}ffler}, {L{\'o}pez}, {Lopez-Lozano}, {Lorenz}, {Loureiro},
  {MacDonald}, {Magalh{\~a}es Fernandes}, {Managau}, {Mann}, {Mantelet},
  {Marchal}, {Marchant}, {Marconi}, {Marie}, {Marinoni}, {Marrese},
  {Marschalk{\'o}}, {Marshall}, {Mart{\'\i}n-Fleitas}, {Martino}, {Mary},
  {Matijevi{\v{c}}}, {Mazeh}, {McMillan}, {Messina}, {Mestre}, {Michalik},
  {Millar}, {Miranda}, {Molina}, {Molinaro}, {Molinaro}, {Moln{\'a}r},
  {Moniez}, {Montegriffo}, {Monteiro}, {Mor}, {Mora}, {Morbidelli}, {Morel},
  {Morgenthaler}, {Morley}, {Morris}, {Mulone}, {Muraveva}, {Musella},
  {Narbonne}, {Nelemans}, {Nicastro}, {Noval}, {Ord{\'e}novic},
  {Ordieres-Mer{\'e}}, {Osborne}, {Pagani}, {Pagano}, {Pailler}, {Palacin},
  {Palaversa}, {Parsons}, {Paulsen}, {Pecoraro}, {Pedrosa}, {Pentik{\"a}inen},
  {Pereira}, {Pichon}, {Piersimoni}, {Pineau}, {Plachy}, {Plum}, {Poujoulet},
  {Pr{\v{s}}a}, {Pulone}, {Ragaini}, {Rago}, {Rambaux}, {Ramos-Lerate},
  {Ranalli}, {Rauw}, {Read}, {Regibo}, {Renk}, {Reyl{\'e}}, {Ribeiro},
  {Rimoldini}, {Ripepi}, {Riva}, {Rixon}, {Roelens}, {Romero-G{\'o}mez},
  {Rowell}, {Royer}, {Rudolph}, {Ruiz-Dern}, {Sadowski}, {Sagrist{\`a}
  Sell{\'e}s}, {Sahlmann}, {Salgado}, {Salguero}, {Sarasso}, {Savietto},
  {Schnorhk}, {Schultheis}, {Sciacca}, {Segol}, {Segovia}, {Segransan},
  {Serpell}, {Shih}, {Smareglia}, {Smart}, {Smith}, {Solano}, {Solitro},
  {Sordo}, {Soria Nieto}, {Souchay}, {Spagna}, {Spoto}, {Stampa}, {Steele},
  {Steidelm{\"u}ller}, {Stephenson}, {Stoev}, {Suess}, {S{\"u}veges}, {Surdej},
  {Szabados}, {Szegedi-Elek}, {Tapiador}, {Taris}, {Tauran}, {Taylor},
  {Teixeira}, {Terrett}, {Tingley}, {Trager}, {Turon}, {Ulla}, {Utrilla},
  {Valentini}, {van Elteren}, {Van Hemelryck}, {van Leeuwen}, {Varadi},
  {Vecchiato}, {Veljanoski}, {Via}, {Vicente}, {Vogt}, {Voss}, {Votruba},
  {Voutsinas}, {Walmsley}, {Weiler}, {Weingrill}, {Werner}, {Wevers},
  {Whitehead}, {Wyrzykowski}, {Yoldas}, {{\v{Z}}erjal}, {Zucker}, {Zurbach},
  {Zwitter}, {Alecu}, {Allen}, {Allende Prieto}, {Amorim},
  {Anglada-Escud{\'e}}, {Arsenijevic}, {Azaz}, {Balm}, {Beck}, {Bernstein},
  {Bigot}, {Bijaoui}, {Blasco}, {Bonfigli}, {Bono}, {Boudreault}, {Bressan},
  {Brown}, {Brunet}, {Bunclark}, {Buonanno}, {Butkevich}, {Carret}, {Carrion},
  {Chemin}, {Ch{\'e}reau}, {Corcione}, {Darmigny}, {de Boer}, {de Teodoro}, {de
  Zeeuw}, {Delle Luche}, {Domingues}, {Dubath}, {Fodor}, {Fr{\'e}zouls},
  {Fries}, {Fustes}, {Fyfe}, {Gallardo}, {Gallegos}, {Gardiol}, {Gebran},
  {Gomboc}, {G{\'o}mez}, {Grux}, {Gueguen}, {Heyrovsky}, {Hoar}, {Iannicola},
  {Isasi Parache}, {Janotto}, {Joliet}, {Jonckheere}, {Keil}, {Kim},
  {Klagyivik}, {Klar}, {Knude}, {Kochukhov}, {Kolka}, {Kos}, {Kutka}, {Lainey},
  {LeBouquin}, {Liu}, {Loreggia}, {Makarov}, {Marseille}, {Martayan},
  {Martinez-Rubi}, {Massart}, {Meynadier}, {Mignot}, {Munari}, {Nguyen},
  {Nordlander}, {Ocvirk}, {O'Flaherty}, {Olias Sanz}, {Ortiz}, {Osorio},
  {Oszkiewicz}, {Ouzounis}, {Palmer}, {Park}, {Pasquato}, {Peltzer}, {Peralta},
  {P{\'e}turaud}, {Pieniluoma}, {Pigozzi}, {Poels}, {Prat}, {Prod'homme},
  {Raison}, {Rebordao}, {Risquez}, {Rocca-Volmerange}, {Rosen}, {Ruiz-Fuertes},
  {Russo}, {Sembay}, {Serraller Vizcaino}, {Short}, {Siebert}, {Silva},
  {Sinachopoulos}, {Slezak}, {Soffel}, {Sosnowska}, {Strai{\v{z}}ys}, {ter
  Linden}, {Terrell}, {Theil}, {Tiede}, {Troisi}, {Tsalmantza}, {Tur},
  {Vaccari}, {Vachier}, {Valles}, {Van Hamme}, {Veltz}, {Virtanen}, {Wallut},
  {Wichmann}, {Wilkinson}, {Ziaeepour}, \& {Zschocke}}]{Gai16}
{Gaia Collaboration}, {Prusti}, T., {de Bruijne}, J.~H.~J., {et~al.} 2016,
  \aap, 595, A1, \dodoi{10.1051/0004-6361/201629272}

\bibitem[{{Gaia Collaboration} {et~al.}(2018){Gaia Collaboration}, {Brown},
  {Vallenari}, {Prusti}, {de Bruijne}, {Babusiaux}, {Bailer-Jones}, {Biermann},
  {Evans}, {Eyer}, {Jansen}, {Jordi}, {Klioner}, {Lammers}, {Lindegren},
  {Luri}, {Mignard}, {Panem}, {Pourbaix}, {Randich}, {Sartoretti}, {Siddiqui},
  {Soubiran}, {van Leeuwen}, {Walton}, {Arenou}, {Bastian}, {Cropper},
  {Drimmel}, {Katz}, {Lattanzi}, {Bakker}, {Cacciari}, {Casta{\~n}eda},
  {Chaoul}, {Cheek}, {De Angeli}, {Fabricius}, {Guerra}, {Holl}, {Masana},
  {Messineo}, {Mowlavi}, {Nienartowicz}, {Panuzzo}, {Portell}, {Riello},
  {Seabroke}, {Tanga}, {Th{\'e}venin}, {Gracia-Abril}, {Comoretto},
  {Garcia-Reinaldos}, {Teyssier}, {Altmann}, {Andrae}, {Audard},
  {Bellas-Velidis}, {Benson}, {Berthier}, {Blomme}, {Burgess}, {Busso},
  {Carry}, {Cellino}, {Clementini}, {Clotet}, {Creevey}, {Davidson}, {De
  Ridder}, {Delchambre}, {Dell'Oro}, {Ducourant},
  {Fern{\'a}ndez-Hern{\'a}ndez}, {Fouesneau}, {Fr{\'e}mat}, {Galluccio},
  {Garc{\'\i}a-Torres}, {Gonz{\'a}lez-N{\'u}{\~n}ez}, {Gonz{\'a}lez-Vidal},
  {Gosset}, {Guy}, {Halbwachs}, {Hambly}, {Harrison}, {Hern{\'a}ndez},
  {Hestroffer}, {Hodgkin}, {Hutton}, {Jasniewicz}, {Jean-Antoine-Piccolo},
  {Jordan}, {Korn}, {Krone-Martins}, {Lanzafame}, {Lebzelter}, {L{\"o}ffler},
  {Manteiga}, {Marrese}, {Mart{\'\i}n-Fleitas}, {Moitinho}, {Mora}, {Muinonen},
  {Osinde}, {Pancino}, {Pauwels}, {Petit}, {Recio-Blanco}, {Richards},
  {Rimoldini}, {Robin}, {Sarro}, {Siopis}, {Smith}, {Sozzetti}, {S{\"u}veges},
  {Torra}, {van Reeven}, {Abbas}, {Abreu Aramburu}, {Accart}, {Aerts},
  {Altavilla}, {{\'A}lvarez}, {Alvarez}, {Alves}, {Anderson}, {Andrei},
  {Anglada Varela}, {Antiche}, {Antoja}, {Arcay}, {Astraatmadja}, {Bach},
  {Baker}, {Balaguer-N{\'u}{\~n}ez}, {Balm}, {Barache}, {Barata}, {Barbato},
  {Barblan}, {Barklem}, {Barrado}, {Barros}, {Barstow}, {Bartholom{\'e}
  Mu{\~n}oz}, {Bassilana}, {Becciani}, {Bellazzini}, {Berihuete}, {Bertone},
  {Bianchi}, {Bienaym{\'e}}, {Blanco-Cuaresma}, {Boch}, {Boeche}, {Bombrun},
  {Borrachero}, {Bossini}, {Bouquillon}, {Bourda}, {Bragaglia}, {Bramante},
  {Breddels}, {Bressan}, {Brouillet}, {Br{\"u}semeister}, {Brugaletta},
  {Bucciarelli}, {Burlacu}, {Busonero}, {Butkevich}, {Buzzi}, {Caffau},
  {Cancelliere}, {Cannizzaro}, {Cantat-Gaudin}, {Carballo}, {Carlucci},
  {Carrasco}, {Casamiquela}, {Castellani}, {Castro-Ginard}, {Charlot},
  {Chemin}, {Chiavassa}, {Cocozza}, {Costigan}, {Cowell}, {Crifo}, {Crosta},
  {Crowley}, {Cuypers}, {Dafonte}, {Damerdji}, {Dapergolas}, {David}, {David},
  {de Laverny}, {De Luise}, {De March}, {de Martino}, {de Souza}, {de Torres},
  {Debosscher}, {del Pozo}, {Delbo}, {Delgado}, {Delgado}, {Di Matteo},
  {Diakite}, {Diener}, {Distefano}, {Dolding}, {Drazinos}, {Dur{\'a}n},
  {Edvardsson}, {Enke}, {Eriksson}, {Esquej}, {Eynard Bontemps}, {Fabre},
  {Fabrizio}, {Faigler}, {Falc{\~a}o}, {Farr{\`a}s Casas}, {Federici},
  {Fedorets}, {Fernique}, {Figueras}, {Filippi}, {Findeisen}, {Fonti},
  {Fraile}, {Fraser}, {Fr{\'e}zouls}, {Gai}, {Galleti}, {Garabato},
  {Garc{\'\i}a-Sedano}, {Garofalo}, {Garralda}, {Gavel}, {Gavras}, {Gerssen},
  {Geyer}, {Giacobbe}, {Gilmore}, {Girona}, {Giuffrida}, {Glass}, {Gomes},
  {Granvik}, {Gueguen}, {Guerrier}, {Guiraud}, {Guti{\'e}rrez-S{\'a}nchez},
  {Haigron}, {Hatzidimitriou}, {Hauser}, {Haywood}, {Heiter}, {Helmi}, {Heu},
  {Hilger}, {Hobbs}, {Hofmann}, {Holland}, {Huckle}, {Hypki}, {Icardi},
  {Jan{\ss}en}, {Jevardat de Fombelle}, {Jonker}, {Juh{\'a}sz}, {Julbe},
  {Karampelas}, {Kewley}, {Klar}, {Kochoska}, {Kohley}, {Kolenberg},
  {Kontizas}, {Kontizas}, {Koposov}, {Kordopatis}, {Kostrzewa-Rutkowska},
  {Koubsky}, {Lambert}, {Lanza}, {Lasne}, {Lavigne}, {Le Fustec}, {Le
  Poncin-Lafitte}, {Lebreton}, {Leccia}, {Leclerc}, {Lecoeur-Taibi},
  {Lenhardt}, {Leroux}, {Liao}, {Licata}, {Lindstr{\o}m}, {Lister}, {Livanou},
  {Lobel}, {L{\'o}pez}, {Managau}, {Mann}, {Mantelet}, {Marchal}, {Marchant},
  {Marconi}, {Marinoni}, {Marschalk{\'o}}, {Marshall}, {Martino}, {Marton},
  {Mary}, {Massari}, {Matijevi{\v{c}}}, {Mazeh}, {McMillan}, {Messina},
  {Michalik}, {Millar}, {Molina}, {Molinaro}, {Moln{\'a}r}, {Montegriffo},
  {Mor}, {Morbidelli}, {Morel}, {Morris}, {Mulone}, {Muraveva}, {Musella},
  {Nelemans}, {Nicastro}, {Noval}, {O'Mullane}, {Ord{\'e}novic},
  {Ord{\'o}{\~n}ez-Blanco}, {Osborne}, {Pagani}, {Pagano}, {Pailler},
  {Palacin}, {Palaversa}, {Panahi}, {Pawlak}, {Piersimoni}, {Pineau}, {Plachy},
  {Plum}, {Poggio}, {Poujoulet}, {Pr{\v{s}}a}, {Pulone}, {Racero}, {Ragaini},
  {Rambaux}, {Ramos-Lerate}, {Regibo}, {Reyl{\'e}}, {Riclet}, {Ripepi}, {Riva},
  {Rivard}, {Rixon}, {Roegiers}, {Roelens}, {Romero-G{\'o}mez}, {Rowell},
  {Royer}, {Ruiz-Dern}, {Sadowski}, {Sagrist{\`a} Sell{\'e}s}, {Sahlmann},
  {Salgado}, {Salguero}, {Sanna}, {Santana-Ros}, {Sarasso}, {Savietto},
  {Schultheis}, {Sciacca}, {Segol}, {Segovia}, {S{\'e}gransan}, {Shih},
  {Siltala}, {Silva}, {Smart}, {Smith}, {Solano}, {Solitro}, {Sordo}, {Soria
  Nieto}, {Souchay}, {Spagna}, {Spoto}, {Stampa}, {Steele},
  {Steidelm{\"u}ller}, {Stephenson}, {Stoev}, {Suess}, {Surdej}, {Szabados},
  {Szegedi-Elek}, {Tapiador}, {Taris}, {Tauran}, {Taylor}, {Teixeira},
  {Terrett}, {Teyssandier}, {Thuillot}, {Titarenko}, {Torra Clotet}, {Turon},
  {Ulla}, {Utrilla}, {Uzzi}, {Vaillant}, {Valentini}, {Valette}, {van Elteren},
  {Van Hemelryck}, {van Leeuwen}, {Vaschetto}, {Vecchiato}, {Veljanoski},
  {Viala}, {Vicente}, {Vogt}, {von Essen}, {Voss}, {Votruba}, {Voutsinas},
  {Walmsley}, {Weiler}, {Wertz}, {Wevers}, {Wyrzykowski}, {Yoldas},
  {{\v{Z}}erjal}, {Ziaeepour}, {Zorec}, {Zschocke}, {Zucker}, {Zurbach}, \&
  {Zwitter}}]{Gai18}
{Gaia Collaboration}, {Brown}, A.~G.~A., {Vallenari}, A., {et~al.} 2018, \aap,
  616, A1, \dodoi{10.1051/0004-6361/201833051}

\bibitem[{{Goliasch} \& {Nelson}(2015)}]{Gol15}
{Goliasch}, J., \& {Nelson}, L. 2015, \apj, 809, 80,
  \dodoi{10.1088/0004-637X/809/1/80}

\bibitem[{{Green} {et~al.}(2019){Green}, {Schlafly}, {Zucker}, {Speagle}, \&
  {Finkbeiner}}]{Gre19}
{Green}, G.~M., {Schlafly}, E., {Zucker}, C., {Speagle}, J.~S., \&
  {Finkbeiner}, D. 2019, \apj, 887, 93, \dodoi{10.3847/1538-4357/ab5362}

\bibitem[{{Green} {et~al.}(2020){Green}, {Marsh}, {Carter}, {Steeghs},
  {Breedt}, {Dhillon}, {Littlefair}, {Parsons}, {Kerry}, {Gentile Fusillo},
  {Ashley}, {Bours}, {Cunningham}, {Dyer}, {G{\"a}nsicke}, {Izquierdo}, {Pala},
  {Pattama}, {Outmani}, {Sahman}, {Sukaum}, \& {Wild}}]{Gre20}
{Green}, M.~J., {Marsh}, T.~R., {Carter}, P.~J., {et~al.} 2020, \mnras, 496,
  1243, \dodoi{10.1093/mnras/staa1509}

\bibitem[{{Hameury} \& {Lasota}(2014)}]{Ham14}
{Hameury}, J.~M., \& {Lasota}, J.~P. 2014, \aap, 569, A48,
  \dodoi{10.1051/0004-6361/201424535}

\bibitem[{{Han} \& {Podsiadlowski}(2004)}]{Han04}
{Han}, Z., \& {Podsiadlowski}, P. 2004, \mnras, 350, 1301,
  \dodoi{10.1111/j.1365-2966.2004.07713.x}

\bibitem[{{Kato}(2019)}]{Kat19c}
{Kato}, T. 2019, \pasj, 71, 20, \dodoi{10.1093/pasj/psy138}

\bibitem[{{Kato} \& {Kojiguchi}(2021)}]{Kat21a}
{Kato}, T., \& {Kojiguchi}, N. 2021, arXiv e-prints, arXiv:2107.14400,
  \dodoi{10.48550/arXiv.2107.14400}

\bibitem[{{Kato} {et~al.}(2019){Kato}, {Pavlenko}, {Pit}, {Antonyuk},
  {Antonyuk}, {Babina}, {Baklanov}, {Sosnovskij}, {Belan}, {Maeda}, {Sugiura},
  {Sumiya}, {Matsumoto}, {Ito}, {Nikai}, {Kojiguchi}, {Matsumoto}, {Dubovsky},
  {Kudzej}, {Medulka}, {Wakamatsu}, {Ohnishi}, {Seki}, {Isogai}, {Simon},
  {Romanjuk}, {Baransky}, {Sergeev}, {Godunova}, {Izviekova}, {Kozlov},
  {Sklyanov}, {Zhuchkov}, {Gutaev}, {Ponomarenko}, {Vasylenko}, {Miller},
  {Kasai}, {Dvorak}, {Menzies}, {de Miguel}, {Brincat}, \& {Pickard}}]{Kat19a}
{Kato}, T., {Pavlenko}, E.~P., {Pit}, N.~V., {et~al.} 2019, \pasj, 71, L1,
  \dodoi{10.1093/pasj/psz007}

\bibitem[{{Kato} {et~al.}(2021){Kato}, {Tampo}, {Kojiguchi}, {Shibata}, {Ito},
  {Isogai}, {Itoh}, {Hambsch}, {Monard}, {Kiyota}, {Vanmunster}, {Sosnovskij},
  {Pavlenko}, {Dubovsky}, {Kudzej}, \& {Medulka}}]{Kat21b}
{Kato}, T., {Tampo}, Y., {Kojiguchi}, N., {et~al.} 2021, \pasj, 73, 1280,
  \dodoi{10.1093/pasj/psab074}

\bibitem[{{Kennedy} {et~al.}(2015){Kennedy}, {Garnavich}, {Callanan}, {Szkody},
  {Littlefield}, \& {Pogge}}]{Ken15}
{Kennedy}, M., {Garnavich}, P., {Callanan}, P., {et~al.} 2015, \apj, 815, 131,
  \dodoi{10.1088/0004-637X/815/2/131}

\bibitem[{{Kim} {et~al.}(2016){Kim}, {Lee}, {Park}, {Kim}, {Cha}, {Lee}, {Han},
  {Chun}, \& {Yuk}}]{Kim16}
{Kim}, S.-L., {Lee}, C.-U., {Park}, B.-G., {et~al.} 2016, Journal of Korean
  Astronomical Society, 49, 37, \dodoi{10.5303/JKAS.2016.49.1.037}

\bibitem[{{Knigge} {et~al.}(2011){Knigge}, {Baraffe}, \& {Patterson}}]{Kni11}
{Knigge}, C., {Baraffe}, I., \& {Patterson}, J. 2011, \apjs, 194, 28,
  \dodoi{10.1088/0067-0049/194/2/28}

\bibitem[{{Lee} {et~al.}(2022){Lee}, {Kim}, {Moon}, {Park}, {Drout}, {Ni}, \&
  {Im}}]{Lee22}
{Lee}, Y., {Kim}, S.~C., {Moon}, D.-S., {et~al.} 2022, \apjl, 925, L22,
  \dodoi{10.3847/2041-8213/ac4c41}

\bibitem[{{Lee} {et~al.}(2019){Lee}, {Moon}, {Kim}, {Park}, {Cha}, \&
  {Lee}}]{Lee19}
{Lee}, Y., {Moon}, D.-S., {Kim}, S.~C., {et~al.} 2019, \apj, 880, 109,
  \dodoi{10.3847/1538-4357/ab2985}

\bibitem[{{Meyer} \& {Meyer-Hofmeister}(1983)}]{Mey83}
{Meyer}, F., \& {Meyer-Hofmeister}, E. 1983, \aap, 121, 29

\bibitem[{{Moon} {et~al.}(2016){Moon}, {Kim}, {Lee}, {Pak}, {Park}, {He},
  {Antoniadis}, {Ni}, {Lee}, {Kim}, {Park}, {Kim}, {Cha}, {Lee}, \&
  {Gonzalez}}]{Moo16}
{Moon}, D.-S., {Kim}, S.~C., {Lee}, J.-J., {et~al.} 2016, in \procspie, Vol.
  9906, Ground-based and Airborne Telescopes VI, 99064I,
  \dodoi{10.1117/12.2233921}

\bibitem[{{Moon} {et~al.}(2021){Moon}, {Ni}, {Drout},
  {Gonz{\'a}lez-Gait{\'a}n}, {Afsariardchi}, {Park}, {Lee}, {Kim},
  {Antoniadis}, {Kim}, \& {Lee}}]{Moo21}
{Moon}, D.-S., {Ni}, Y.~Q., {Drout}, M.~R., {et~al.} 2021, \apj, 910, 151,
  \dodoi{10.3847/1538-4357/abe466}

\bibitem[{{Ni} {et~al.}(2023{\natexlab{a}}){Ni}, {Moon}, {Drout}, {Matzner},
  {Leong}, {Kim}, {Park}, \& {Lee}}]{Ni23b}
{Ni}, Y.~Q., {Moon}, D.-S., {Drout}, M.~R., {et~al.} 2023{\natexlab{a}}, arXiv
  e-prints, arXiv:2304.00625, \dodoi{10.48550/arXiv.2304.00625}

\bibitem[{{Ni} {et~al.}(2022){Ni}, {Moon}, {Drout}, {Polin}, {Sand},
  {Gonz{\'a}lez-Gait{\'a}n}, {Kim}, {Lee}, {Park}, {Howell}, {Nugent}, {Piro},
  {Brown}, {Galbany}, {Burke}, {Hiramatsu}, {Hosseinzadeh}, {Valenti},
  {Afsariardchi}, {Andrews}, {Antoniadis}, {Arcavi}, {Beaton}, {Bostroem},
  {Carlberg}, {Cenko}, {Cha}, {Dong}, {Gal-Yam}, {Haislip}, {Holoien},
  {Johnson}, {Kouprianov}, {Lee}, {Matzner}, {Morrell}, {McCully}, {Pignata},
  {Reichart}, {Rich}, {Ryder}, {Smith}, {Wyatt}, \& {Yang}}]{Ni22}
---. 2022, Nature Astronomy, 6, 568, \dodoi{10.1038/s41550-022-01603-4}

\bibitem[{{Ni} {et~al.}(2023{\natexlab{b}}){Ni}, {Moon}, {Drout}, {Polin},
  {Sand}, {Gonz{\'a}lez-Gait{\'a}n}, {Kim}, {Lee}, {Park}, {Howell}, {Nugent},
  {Piro}, {Brown}, {Galbany}, {Burke}, {Hiramatsu}, {Hosseinzadeh}, {Valenti},
  {Afsariardchi}, {Andrews}, {Antoniadis}, {Beaton}, {Bostroem}, {Carlberg},
  {Cenko}, {Cha}, {Dong}, {Gal-Yam}, {Haislip}, {Holoien}, {Johnson},
  {Kouprianov}, {Lee}, {Matzner}, {Morrell}, {McCully}, {Pignata}, {Reichart},
  {Rich}, {Ryder}, {Smith}, {Wyatt}, \& {Yang}}]{Ni23}
---. 2023{\natexlab{b}}, \apj, 946, 7, \dodoi{10.3847/1538-4357/aca9be}

\bibitem[{{Ohshima}(2023)}]{Ohs23}
{Ohshima}, T. 2023, arXiv e-prints, arXiv:2301.08579,
  \dodoi{10.48550/arXiv.2301.08579}

\bibitem[{{Olech}(1997)}]{Ole97}
{Olech}, A. 1997, \actaa, 47, 281.
\newblock \doarXiv{astro-ph/9706180}

\bibitem[{{Osaki}(1996)}]{Osa96}
{Osaki}, Y. 1996, \pasp, 108, 39, \dodoi{10.1086/133689}

\bibitem[{{Osterbrock}(1989)}]{Ost89}
{Osterbrock}, D.~E. 1989, {Astrophysics of gaseous nebulae and active galactic
  nuclei}

\bibitem[{{Otulakowska-Hypka} {et~al.}(2016){Otulakowska-Hypka}, {Olech}, \&
  {Patterson}}]{Otu16}
{Otulakowska-Hypka}, M., {Olech}, A., \& {Patterson}, J. 2016, \mnras, 460,
  2526, \dodoi{10.1093/mnras/stw1120}

\bibitem[{{Paczynski} \& {Schwarzenberg-Czerny}(1980)}]{Pac80}
{Paczynski}, B., \& {Schwarzenberg-Czerny}, A. 1980, \actaa, 30, 127

\bibitem[{{Paczynski} \& {Sienkiewicz}(1981)}]{Pac81}
{Paczynski}, B., \& {Sienkiewicz}, R. 1981, \apjl, 248, L27,
  \dodoi{10.1086/183616}

\bibitem[{{Park} {et~al.}(2017){Park}, {Moon}, {Zaritsky}, {Pak}, {Lee}, {Kim},
  {Kim}, \& {Cha}}]{Par17}
{Park}, H.~S., {Moon}, D.-S., {Zaritsky}, D., {et~al.} 2017, \apj, 848, 19,
  \dodoi{10.3847/1538-4357/aa88ab}

\bibitem[{{Patterson}(2011)}]{Pat11}
{Patterson}, J. 2011, \mnras, 411, 2695,
  \dodoi{10.1111/j.1365-2966.2010.17881.x}

\bibitem[{{Patterson} {et~al.}(2008){Patterson}, {Thorstensen}, \&
  {Knigge}}]{Pat08}
{Patterson}, J., {Thorstensen}, J.~R., \& {Knigge}, C. 2008, \pasp, 120, 510,
  \dodoi{10.1086/588615}

\bibitem[{{Paxton} {et~al.}(2011){Paxton}, {Bildsten}, {Dotter}, {Herwig},
  {Lesaffre}, \& {Timmes}}]{Pax11}
{Paxton}, B., {Bildsten}, L., {Dotter}, A., {et~al.} 2011, \apjs, 192, 3,
  \dodoi{10.1088/0067-0049/192/1/3}

\bibitem[{{Paxton} {et~al.}(2013){Paxton}, {Cantiello}, {Arras}, {Bildsten},
  {Brown}, {Dotter}, {Mankovich}, {Montgomery}, {Stello}, {Timmes}, \&
  {Townsend}}]{Pax13}
{Paxton}, B., {Cantiello}, M., {Arras}, P., {et~al.} 2013, \apjs, 208, 4,
  \dodoi{10.1088/0067-0049/208/1/4}

\bibitem[{{Paxton} {et~al.}(2015){Paxton}, {Marchant}, {Schwab}, {Bauer},
  {Bildsten}, {Cantiello}, {Dessart}, {Farmer}, {Hu}, {Langer}, {Townsend},
  {Townsley}, \& {Timmes}}]{Pax15}
{Paxton}, B., {Marchant}, P., {Schwab}, J., {et~al.} 2015, \apjs, 220, 15,
  \dodoi{10.1088/0067-0049/220/1/15}

\bibitem[{{Paxton} {et~al.}(2018){Paxton}, {Schwab}, {Bauer}, {Bildsten},
  {Blinnikov}, {Duffell}, {Farmer}, {Goldberg}, {Marchant}, {Sorokina},
  {Thoul}, {Townsend}, \& {Timmes}}]{Pax18}
{Paxton}, B., {Schwab}, J., {Bauer}, E.~B., {et~al.} 2018, \apjs, 234, 34,
  \dodoi{10.3847/1538-4365/aaa5a8}

\bibitem[{{Paxton} {et~al.}(2019){Paxton}, {Smolec}, {Schwab}, {Gautschy},
  {Bildsten}, {Cantiello}, {Dotter}, {Farmer}, {Goldberg}, {Jermyn}, {Kanbur},
  {Marchant}, {Thoul}, {Townsend}, {Wolf}, {Zhang}, \& {Timmes}}]{Pax19}
{Paxton}, B., {Smolec}, R., {Schwab}, J., {et~al.} 2019, \apjs, 243, 10,
  \dodoi{10.3847/1538-4365/ab2241}

\bibitem[{{Pichardo Marcano} {et~al.}(2021){Pichardo Marcano}, {Rivera
  Sandoval}, {Maccarone}, \& {Scaringi}}]{Pic21}
{Pichardo Marcano}, M., {Rivera Sandoval}, L.~E., {Maccarone}, T.~J., \&
  {Scaringi}, S. 2021, \mnras, 508, 3275, \dodoi{10.1093/mnras/stab2685}

\bibitem[{{Podsiadlowski} {et~al.}(2003){Podsiadlowski}, {Han}, \&
  {Rappaport}}]{Pod03}
{Podsiadlowski}, P., {Han}, Z., \& {Rappaport}, S. 2003, \mnras, 340, 1214,
  \dodoi{10.1046/j.1365-8711.2003.06380.x}

\bibitem[{{Rappaport} {et~al.}(1982){Rappaport}, {Joss}, \& {Webbink}}]{Rap82}
{Rappaport}, S., {Joss}, P.~C., \& {Webbink}, R.~F. 1982, \apj, 254, 616,
  \dodoi{10.1086/159772}

\bibitem[{{Rappaport} {et~al.}(1983){Rappaport}, {Verbunt}, \& {Joss}}]{Rap83}
{Rappaport}, S., {Verbunt}, F., \& {Joss}, P.~C. 1983, \apj, 275, 713,
  \dodoi{10.1086/161569}

\bibitem[{{Ritter} \& {Kolb}(2003)}]{Rit03}
{Ritter}, H., \& {Kolb}, U. 2003, \aap, 404, 301,
  \dodoi{10.1051/0004-6361:20030330}

\bibitem[{{S{\'a}nchez-Bl{\'a}zquez} {et~al.}(2006){S{\'a}nchez-Bl{\'a}zquez},
  {Peletier}, {Jim{\'e}nez-Vicente}, {Cardiel}, {Cenarro},
  {Falc{\'o}n-Barroso}, {Gorgas}, {Selam}, \& {Vazdekis}}]{San06}
{S{\'a}nchez-Bl{\'a}zquez}, P., {Peletier}, R.~F., {Jim{\'e}nez-Vicente}, J.,
  {et~al.} 2006, \mnras, 371, 703, \dodoi{10.1111/j.1365-2966.2006.10699.x}

\bibitem[{{Simonsen}(2011)}]{Sim11}
{Simonsen}, M. 2011, \jaavso, 39, 66, \dodoi{10.48550/arXiv.1104.0967}

\bibitem[{{Smak}(2000)}]{Sma00}
{Smak}, J. 2000, \nar, 44, 171, \dodoi{10.1016/S1387-6473(00)00033-6}

\bibitem[{{Stetson}(1987)}]{Ste87}
{Stetson}, P.~B. 1987, \pasp, 99, 191, \dodoi{10.1086/131977}

\bibitem[{{Szkody} {et~al.}(2013){Szkody}, {Albright}, {Linnell}, {Everett},
  {McMillan}, {Saurage}, {Huehnerhoff}, {Howell}, {Simonsen}, \&
  {Hunt-Walker}}]{Szk13}
{Szkody}, P., {Albright}, M., {Linnell}, A.~P., {et~al.} 2013, \pasp, 125,
  1421, \dodoi{10.1086/674170}

\bibitem[{{Thorstensen} {et~al.}(2002){Thorstensen}, {Fenton}, {Patterson},
  {Kemp}, {Krajci}, \& {Baraffe}}]{Tho02a}
{Thorstensen}, J.~R., {Fenton}, W.~H., {Patterson}, J.~O., {et~al.} 2002,
  \apjl, 567, L49, \dodoi{10.1086/339905}

\bibitem[{{Warner}(1995)}]{War95}
{Warner}, B. 1995, Cambridge Astrophysics Series, 28

\bibitem[{{Wright} {et~al.}(2010){Wright}, {Eisenhardt}, {Mainzer}, {Ressler},
  {Cutri}, {Jarrett}, {Kirkpatrick}, {Padgett}, {McMillan}, {Skrutskie},
  {Stanford}, {Cohen}, {Walker}, {Mather}, {Leisawitz}, {Gautier}, {McLean},
  {Benford}, {Lonsdale}, {Blain}, {Mendez}, {Irace}, {Duval}, {Liu}, {Royer},
  {Heinrichsen}, {Howard}, {Shannon}, {Kendall}, {Walsh}, {Larsen}, {Cardon},
  {Schick}, {Schwalm}, {Abid}, {Fabinsky}, {Naes}, \& {Tsai}}]{Wri10}
{Wright}, E.~L., {Eisenhardt}, P. R.~M., {Mainzer}, A.~K., {et~al.} 2010, \aj,
  140, 1868, \dodoi{10.1088/0004-6256/140/6/1868}

\end{thebibliography}
\end{document}